\pdfoutput=1
\documentclass[12pt,a4paper]{article}

\usepackage{ifthen} % for conditional statements
\usepackage[bottom]{footmisc}
\newboolean{pdflatex}
\setboolean{pdflatex}{true} % False for eps figures 

\newboolean{articletitles}
\setboolean{articletitles}{true} % False removes titles in references

\newboolean{uprightparticles}
\setboolean{uprightparticles}{false} %True for upright particle symbols

\newboolean{inbibliography}
\setboolean{inbibliography}{false} %True once you enter the bibliography

\def\paperauthors{LHCb collaboration} % Leave as is for PAPER and CONF
\def\paperasciititle{Search for Bc decays to two charm mesons} % Set ASCII title here
\def\papertitle{Search for \Bc decays to two charm mesons} % Latex formatted title
\def\paperkeywords{{High Energy Physics}, {LHCb}} % Comma separated list
\def\papercopyright{CERN on behalf of the LHCb collaboration}
\def\paperlicence{CC-BY-4.0}
\def\paperlicenceurl{https://creativecommons.org/licenses/by/4.0/}

\usepackage[top=1in, bottom=1.25in, left=1in, right=1in]{geometry}

\usepackage{microtype}
\usepackage{lineno}  % for line numbering during review
\usepackage{xspace} % To avoid problems with missing or double spaces after
\usepackage{caption} %these three command get the figure and table captions automatically small

\usepackage{graphicx}  % to include figures (can also use other packages)
\usepackage{color}
\usepackage{colortbl}
\graphicspath{{./figs/}} % Make Latex search fig subdir for figures

\usepackage{amsmath} % Adds a large collection of math symbols
\usepackage{amssymb}
\usepackage{amsfonts}
\usepackage{upgreek} % Adds in support for greek letters in roman typeset

\newcommand*\patchAmsMathEnvironmentForLineno[1]{%
\expandafter\let\csname old#1\expandafter\endcsname\csname #1\endcsname
\expandafter\let\csname oldend#1\expandafter\endcsname\csname
end#1\endcsname
 \renewenvironment{#1}%
   {\linenomath\csname old#1\endcsname}%
   {\csname oldend#1\endcsname\endlinenomath}%
}
\newcommand*\patchBothAmsMathEnvironmentsForLineno[1]{%
  \patchAmsMathEnvironmentForLineno{#1}%
  \patchAmsMathEnvironmentForLineno{#1*}%
}
\AtBeginDocument{%
\patchBothAmsMathEnvironmentsForLineno{equation}%
\patchBothAmsMathEnvironmentsForLineno{align}%
\patchBothAmsMathEnvironmentsForLineno{flalign}%
\patchBothAmsMathEnvironmentsForLineno{alignat}%
\patchBothAmsMathEnvironmentsForLineno{gather}%
\patchBothAmsMathEnvironmentsForLineno{multline}%
\patchBothAmsMathEnvironmentsForLineno{eqnarray}%
}

\usepackage{hyperxmp}

\usepackage[pdftex,
            pdfauthor={\paperauthors},
            pdftitle={\paperasciititle},
            pdfkeywords={\paperkeywords},
            pdfcopyright={Copyright (C) \papercopyright},
            pdflicenseurl={\paperlicenceurl}]{hyperref}

\usepackage[all]{hypcap} % Internal hyperlinks to floats.

\usepackage{xspace} 
\usepackage{upgreek}

\def\lhcb {\mbox{LHCb}\xspace}

\def\MagUp {\mbox{\em Mag\kern -0.05em Up}\xspace}

\ifthenelse{\boolean{uprightparticles}}%
{

 \def\Ppi         {\ensuremath{\uppi}\xspace}

 \def\Ppsi        {\ensuremath{\uppsi}\xspace}

 \def\PDelta      {\ensuremath{\Delta}\xspace}                 
 \def\PXi      {\ensuremath{\Xi}\xspace}                 
 \def\PLambda      {\ensuremath{\Lambda}\xspace}                 
 \def\PSigma      {\ensuremath{\Sigma}\xspace}                 
 \def\POmega      {\ensuremath{\Omega}\xspace}                 
 \def\PUpsilon      {\ensuremath{\Upsilon}\xspace}                 
 
 \def\PB      {\ensuremath{\mathrm{B}}\xspace}                 
                  
 \def\PD      {\ensuremath{\mathrm{D}}\xspace}

 \def\PJ      {\ensuremath{\mathrm{J}}\xspace}                 
 \def\PK      {\ensuremath{\mathrm{K}}\xspace}

 \def\Pb      {\ensuremath{\mathrm{b}}\xspace}                 
 \def\Pc      {\ensuremath{\mathrm{c}}\xspace}                 
 \def\Pd      {\ensuremath{\mathrm{d}}\xspace}

 \def\Pi      {\ensuremath{\mathrm{i}}\xspace}

 \def\Ps      {\ensuremath{\mathrm{s}}\xspace}                 
                  
 \def\Pu      {\ensuremath{\mathrm{u}}\xspace}

}
{

 \def\Ppi         {\ensuremath{\pi}\xspace}

 \def\Ppsi        {\ensuremath{\psi}\xspace}                 
                  
 \mathchardef\PDelta="7101
 \mathchardef\PXi="7104
 \mathchardef\PLambda="7103
 \mathchardef\PSigma="7106
 \mathchardef\POmega="710A
 \mathchardef\PUpsilon="7107
                  
 \def\PB      {\ensuremath{B}\xspace}                 
                  
 \def\PD      {\ensuremath{D}\xspace}

 \def\PJ      {\ensuremath{J}\xspace}                 
 \def\PK      {\ensuremath{K}\xspace}

 \def\Pb      {\ensuremath{b}\xspace}                 
 \def\Pc      {\ensuremath{c}\xspace}                 
 \def\Pd      {\ensuremath{d}\xspace}

 \def\Pi      {\ensuremath{i}\xspace}

 \def\Ps      {\ensuremath{s}\xspace}                 
                  
 \def\Pu      {\ensuremath{u}\xspace}

}

\makeatletter
\ifcase \@ptsize \relax% 10pt
  \newcommand{\miniscule}{\@setfontsize\miniscule{4}{5}}% \tiny: 5/6
\or% 11pt
  \newcommand{\miniscule}{\@setfontsize\miniscule{5}{6}}% \tiny: 6/7
\or% 12pt
  \newcommand{\miniscule}{\@setfontsize\miniscule{5}{6}}% \tiny: 6/7
\fi
\makeatother

\DeclareRobustCommand{\optbar}[1]{\shortstack{{\miniscule (\rule[.5ex]{1.25em}{.18mm})}
  \\ [-.7ex] $#1$}}

\def\uquark    {{\ensuremath{\Pu}}\xspace}

\def\dquark    {{\ensuremath{\Pd}}\xspace}

\def\squark    {{\ensuremath{\Ps}}\xspace}

\def\cquark    {{\ensuremath{\Pc}}\xspace}

\def\bquark    {{\ensuremath{\Pb}}\xspace}

\def\pion   {{\ensuremath{\Ppi}}\xspace}
\def\piz    {{\ensuremath{\pion^0}}\xspace}

\def\pip    {{\ensuremath{\pion^+}}\xspace}
\def\pim    {{\ensuremath{\pion^-}}\xspace}

\def\kaon    {{\ensuremath{\PK}}\xspace}
  \def\Kbar    {{\kern 0.2em\overline{\kern -0.2em \PK}{}}\xspace}

\def\KorKbar    {\kern 0.18em\optbar{\kern -0.18em K}{}\xspace}

\def\Kp      {{\ensuremath{\kaon^+}}\xspace}
\def\Km      {{\ensuremath{\kaon^-}}\xspace}

  \def\Dbar    {{\kern 0.2em\overline{\kern -0.2em \PD}{}}\xspace}
\def\D       {{\ensuremath{\PD}}\xspace}
\def\Db      {{\ensuremath{\Dbar}}\xspace}
\def\DorDbar    {\kern 0.18em\optbar{\kern -0.18em D}{}\xspace}
\def\Dz      {{\ensuremath{\D^0}}\xspace}
\def\Dzb     {{\ensuremath{\Dbar{}^0}}\xspace}
\def\Dp      {{\ensuremath{\D^+}}\xspace}

\def\Dstarz  {{\ensuremath{\D^{*0}}}\xspace}
\def\Dstarzb {{\ensuremath{\Dbar{}^{*0}}}\xspace}
\def\Dstarp  {{\ensuremath{\D^{*+}}}\xspace}

\def\Ds      {{\ensuremath{\D^+_\squark}}\xspace}
\def\Dsp     {{\ensuremath{\D^+_\squark}}\xspace}

\def\Dss     {{\ensuremath{\D^{*+}_\squark}}\xspace}

\def\B       {{\ensuremath{\PB}}\xspace}
\def\Bbar    {{\ensuremath{\kern 0.18em\overline{\kern -0.18em \PB}{}}}\xspace}

\def\BorBbar    {\kern 0.18em\optbar{\kern -0.18em B}{}\xspace}

\def\Bu      {{\ensuremath{\B^+}}\xspace}

\def\Bp      {{\ensuremath{\Bu}}\xspace}

\def\Bc      {{\ensuremath{\B_\cquark^+}}\xspace}

\def\jpsi     {{\ensuremath{{\PJ\mskip -3mu/\mskip -2mu\Ppsi\mskip 2mu}}}\xspace}

  \def\Y#1S{\ensuremath{\PUpsilon{(#1S)}}\xspace}% no space before {...}!

\def\Lbar        {{\ensuremath{\kern 0.1em\overline{\kern -0.1em\PLambda}}}\xspace}
\def\LorLbar    {\kern 0.18em\optbar{\kern -0.18em \PLambda}{}\xspace}

\def\BF         {{\ensuremath{\mathcal{B}}}\xspace}

\newcommand{\decay}[2]{\ensuremath{#1\!\to #2}\xspace}         % {\Pa}{\Pb \Pc}

\def\to                 {\ensuremath{\rightarrow}\xspace}

\def\CP                {{\ensuremath{C\!P}}\xspace}

\def\Vud  {{\ensuremath{V_{\uquark\dquark}}}\xspace}
\def\Vcd  {{\ensuremath{V_{\cquark\dquark}}}\xspace}

\def\Vub  {{\ensuremath{V_{\uquark\bquark}}}\xspace}
\def\Vcb  {{\ensuremath{V_{\cquark\bquark}}}\xspace}

\def\Vubs  {{\ensuremath{V_{\uquark\bquark}^\ast}}\xspace}
\def\Vcbs  {{\ensuremath{V_{\cquark\bquark}^\ast}}\xspace}

\def\AT#1     {\ensuremath{A_{\mathrm{T}}^{#1}}\xspace}           % 2

\def\C#1      {\ensuremath{\mathcal{C}_{#1}}\xspace}                       % 9
\def\Cp#1     {\ensuremath{\mathcal{C}_{#1}^{'}}\xspace}                    % 7
\def\Ceff#1   {\ensuremath{\mathcal{C}_{#1}^{\mathrm{(eff)}}}\xspace}        % 9  
\def\Cpeff#1  {\ensuremath{\mathcal{C}_{#1}^{'\mathrm{(eff)}}}\xspace}       % 7
\def\Ope#1    {\ensuremath{\mathcal{O}_{#1}}\xspace}                       % 2
\def\Opep#1   {\ensuremath{\mathcal{O}_{#1}^{'}}\xspace}                    % 7

\newcommand{\tev}{\ifthenelse{\boolean{inbibliography}}{\ensuremath{~T\kern -0.05em eV}}{\ensuremath{\mathrm{\,Te\kern -0.1em V}}}\xspace}
\newcommand{\gev}{\ensuremath{\mathrm{\,Ge\kern -0.1em V}}\xspace}
\newcommand{\mev}{\ensuremath{\mathrm{\,Me\kern -0.1em V}}\xspace}
\newcommand{\kev}{\ensuremath{\mathrm{\,ke\kern -0.1em V}}\xspace}
\newcommand{\ev}{\ensuremath{\mathrm{\,e\kern -0.1em V}}\xspace}
\newcommand{\gevc}{\ensuremath{{\mathrm{\,Ge\kern -0.1em V\!/}c}}\xspace}
\newcommand{\mevc}{\ensuremath{{\mathrm{\,Me\kern -0.1em V\!/}c}}\xspace}
\newcommand{\gevcc}{\ensuremath{{\mathrm{\,Ge\kern -0.1em V\!/}c^2}}\xspace}
\newcommand{\gevgevcccc}{\ensuremath{{\mathrm{\,Ge\kern -0.1em V^2\!/}c^4}}\xspace}
\newcommand{\mevcc}{\ensuremath{{\mathrm{\,Me\kern -0.1em V\!/}c^2}}\xspace}

\def\mum  {\ensuremath{{\,\upmu\mathrm{m}}}\xspace}

\def\invfb   {\ensuremath{\mbox{\,fb}^{-1}}\xspace}

\newcommand{\chisq}{\ensuremath{\chi^2}\xspace}

\newcommand{\chisqip}{\ensuremath{\chi^2_{\text{IP}}}\xspace}

\def\gsim{{~\raise.15em\hbox{$>$}\kern-.85em
          \lower.35em\hbox{$\sim$}~}\xspace}
\def\lsim{{~\raise.15em\hbox{$<$}\kern-.85em
          \lower.35em\hbox{$\sim$}~}\xspace}

\def\pt         {\mbox{$p_{\mathrm{ T}}$}\xspace}

\def\bcvegpy    {\mbox{\textsc{Bcvegpy}}\xspace}

\def\evtgen     {\mbox{\textsc{EvtGen}}\xspace}

\def\geant      {\mbox{\textsc{Geant4}}\xspace}

\def\photos     {\mbox{\textsc{Photos}}\xspace}

\def\pythia     {\mbox{\textsc{Pythia}}\xspace}

\def\tell1  {TELL1\xspace}
\def\ukl1   {UKL1\xspace}

\usepackage{cite} % Allows for ranges in citations
\usepackage{mciteplus}

\def\DporDs{{\ensuremath{\D^+_{(\squark)}}}\xspace}
\def\BporBc{{\ensuremath{\B^+_{(\cquark)}}}\xspace}
\def\fcfu{{\ensuremath{\frac{f_c}{f_u}}}\xspace}
\def\inlfcfu{{\ensuremath{f_c/f_u}}\xspace}
\def\DKpi{{\ensuremath{\decay{\Dz}{\Km\pip}}}\xspace}

\def\DKpipipi{{\ensuremath{\decay{\Dz}{\Km\pip\pim\pip}}}\xspace}

\def\DzorDzstar{{\ensuremath{\D^{(*)0}}}\xspace}
\def\DzborDzbstar{{\ensuremath{\Db^{(*)0}}}\xspace}
\def\DporDsstar{{\ensuremath{\D^{*+}_{(\squark)}}}\xspace}
\def\DporDsorstar{{\ensuremath{\D^{(*)+}_{(\squark)}}}\xspace}

\def\DzorDzb{{\ensuremath{\DorDbar^0}}\xspace}
\def\DzorDzbstar{{\ensuremath{\DorDbar^{*0}}}\xspace}
\def\DzorDzborstar{{\ensuremath{\DorDbar^{(*)0}}}\xspace}

\begin{document}

%%%%%%%%%%%%%%%%%%%%%%%%%
%%%%% Title     %%%%%%%%%
%%%%%%%%%%%%%%%%%%%%%%%%%
\renewcommand{\thefootnote}{\fnsymbol{footnote}}
\setcounter{footnote}{1}

% %%%%%%% CHOOSE TITLE PAGE--------
%\onecolumn
%\input{title-LHCb-INT}
%\input{title-LHCb-ANA}
%\input{title-LHCb-CONF}
% $Id: title-LHCb-PAPER.tex 111203 2017-08-08 15:28:40Z pkoppenb $
% ===============================================================================
% Purpose: LHCb-PAPER journal paper title page template
% Author: 
% Created on: 2010-09-25
% ===============================================================================

%%%%%%%%%%%%%%%%%%%%%%%%%
%%%%%  TITLE PAGE  %%%%%%
%%%%%%%%%%%%%%%%%%%%%%%%%
\begin{titlepage}
\pagenumbering{roman}

% Header ---------------------------------------------------
\vspace*{-1.5cm}
\centerline{\large EUROPEAN ORGANIZATION FOR NUCLEAR RESEARCH (CERN)}
\vspace*{1.5cm}
\noindent
\begin{tabular*}{\linewidth}{lc@{\extracolsep{\fill}}r@{\extracolsep{0pt}}}
\ifthenelse{\boolean{pdflatex}}% Logo format choice
{\vspace*{-1.5cm}\mbox{\!\!\!\includegraphics[width=.14\textwidth]{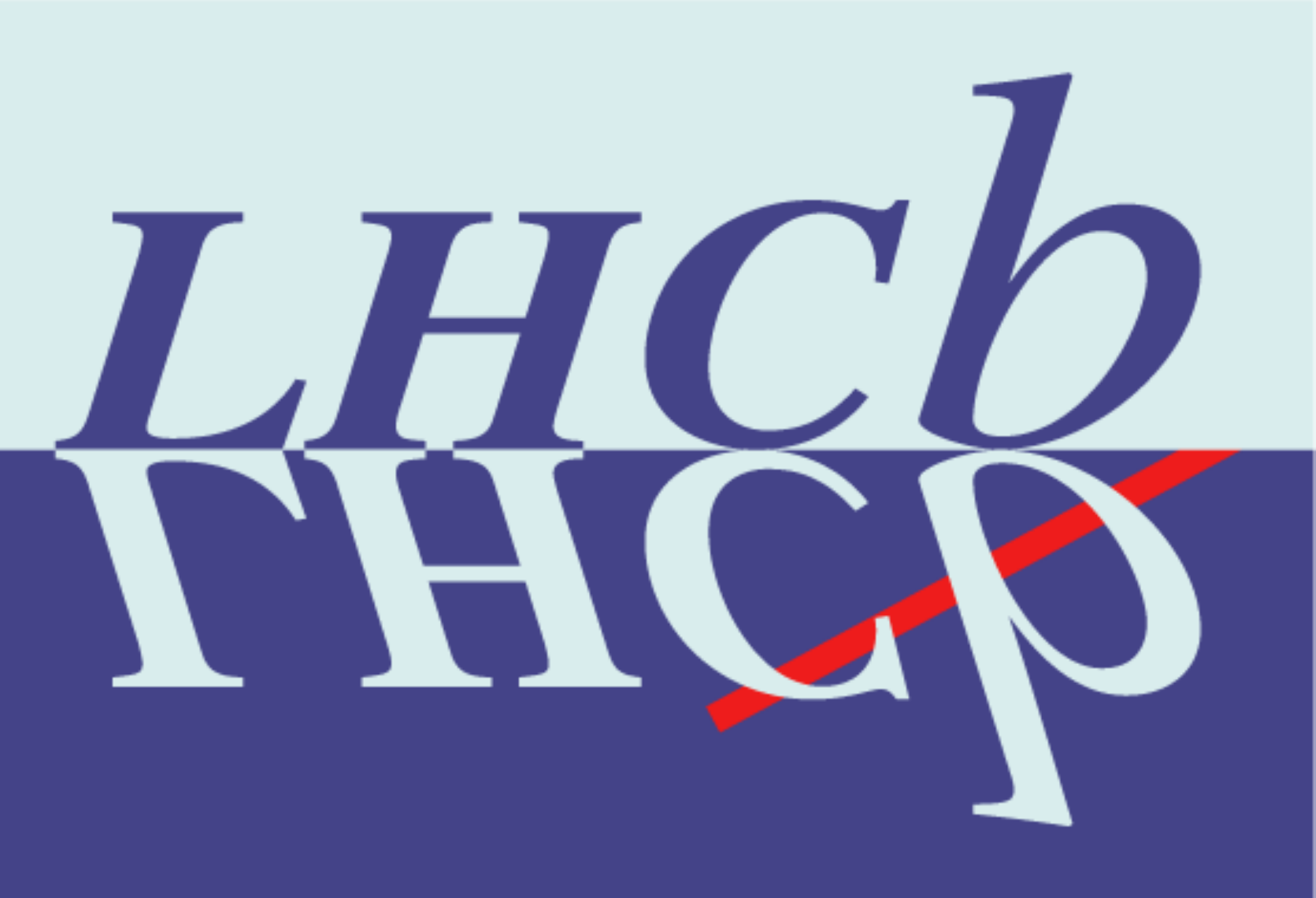}} & &}%
{\vspace*{-1.2cm}\mbox{\!\!\!\includegraphics[width=.12\textwidth]{lhcb-logo.eps}} & &}%
\\
 & & CERN-EP-2017-313 \\  % ID 
 & & LHCb-PAPER-2017-045 \\  % ID 
% & & \today \\ % Date - Can also hardwire e.g.: 23 March 2010
 & & December 13, 2017
% not in paper \hline
\end{tabular*}

\vspace*{4.0cm}

% Title --------------------------------------------------
{\normalfont\bfseries\boldmath\huge
\begin{center}
% DO NOT EDIT HERE. Instead edit macro in main.tex to keep metadata correct
  \papertitle 
\end{center}
}

\vspace*{2.0cm}

% Authors -------------------------------------------------
\begin{center}
%In the footnote, replace 'paper' by 'Letter' in case of submission to PRL or PLB 
% Edit macro in main.tex to keep metadata correct
\paperauthors\footnote{Authors are listed at the end of this paper.}
\end{center}

\vspace{\fill}

% Abstract -----------------------------------------------
\begin{abstract}
  \noindent

A search for decays of $B_c^+$ mesons to two charm mesons is performed for the first time
using data corresponding to an integrated luminosity of 3.0\invfb, collected by the LHCb experiment
in $pp$ collisions at centre-of-mass energies of 7 and 8\,TeV.
The decays considered are \decay{\Bc}{\DporDsorstar\DzborDzbstar} and \decay{\Bc}{\DporDsorstar\DzorDzstar},
which are normalised to high-yield \decay{\Bp}{\DporDs\Dzb} decays.
No evidence for a signal is found and limits are set on twelve \Bc decay modes.
\end{abstract}

\vspace*{2.0cm}

\begin{center}
  Published in Nucl.~Phys.~B 930 (2018) 563-582
\end{center}

\vspace{\fill}

{\footnotesize 
% Edit macro in main.tex to keep metadata correct
\centerline{\copyright~\papercopyright, licence \href{\paperlicenceurl}{\paperlicence}.}}
\vspace*{2mm}

\end{titlepage}

%%%%%%%%%%%%%%%%%%%%%%%%%%%%%%%%
%%%%%  EOD OF TITLE PAGE  %%%%%%
%%%%%%%%%%%%%%%%%%%%%%%%%%%%%%%%

%  empty page follows the title page ----
\newpage
\setcounter{page}{2}
\mbox{~}
%\newpage
%

\cleardoublepage

%\twocolumn
% %%%%%%%%%%%%% ---------

\renewcommand{\thefootnote}{\arabic{footnote}}
\setcounter{footnote}{0}

%%%%%%%%%%%%%%%%%%%%%%%%%
%%%%% Main text %%%%%%%%%
%%%%%%%%%%%%%%%%%%%%%%%%%

\pagestyle{plain} % restore page numbers for the main text
\setcounter{page}{1}
\pagenumbering{arabic}

%% Uncomment during review phase. 
%% Comment before a final submission.
%\linenumbers

\setcounter{figure}{0}
\setcounter{table}{0}

\section{Introduction}
\label{sec:Introduction}

Flavour transitions between quarks are governed 
in the Standard Model (SM) of elementary particle physics by the Cabibbo-Kobayashi-Maskawa (CKM) 
quark-mixing matrix~\cite{Cabibbo:1963yz,Kobayashi:1973fv}.
Here the transition amplitudes between up-type quarks, $q$, and down-type quarks, $q'$, are described by the complex numbers $V_{qq'}$,
defining the $3\times3$ unitary CKM matrix.
Precision measurements of the magnitude and phase of the CKM matrix elements may reveal 
signs of new physics if observables that could be affected by new particles are found to be inconsistent
with SM predictions.

One parameter of particular interest is $\gamma\equiv {\rm arg}(-\Vud\Vubs/\Vcd\Vcbs)$,
which can be determined experimentally with negligible theoretical uncertainties
from the charge-parity (\CP) asymmetry caused by the interference between $\bquark\to\uquark$ and $\bquark\to\cquark$ transitions.
Presently, the most precise determinations of $\gamma$ come from measurements of
the \CP asymmetry in \mbox{\decay{\Bp}{\DzorDzb\Kp}} decays.~\cite{LHCb-PAPER-2017-021,LHCb-PAPER-2016-032}.\footnote{Unless
specified otherwise, charge conjugation is implied throughout the paper.}
%where \DzorDzb stands for a \Dz or a \Dzb meson

Decays of \Bc mesons to two charm mesons, \mbox{\decay{\Bc}{\DporDs\DzorDzb}}, have also been proposed 
to measure $\gamma$~\cite{Masetti:1992in,Fleischer:2000pp,Giri:2001be,Giri:2006cw}.
Decays with one excited charm meson in the final state, \mbox{\decay{\Bc}{\DporDsstar\DzorDzb}}
and \mbox{\decay{\Bc}{\DporDs\DzorDzbstar}}, can be used for measuring the angle $\gamma$  in the same way as \mbox{\decay{\Bc}{\DporDs\D}} decays.
For \Bc decays with two excited charm mesons, \mbox{\decay{\Bc}{\DporDsstar\DzorDzbstar}}, angular distributions provide
an alternative method to determine $\gamma$~\cite{Giri:2001be}.
Some predicted branching fractions are listed in Table~\ref{table:bfestimate}. 

\begin{table}[b]
  \caption{
    Estimates of the branching fractions of four \mbox{\decay{\Bc}{\DporDs\DzorDzb}} decays in units of $10^{-6}$.
    Decays of the \Bc meson to final states with one or two excited charm mesons have similar branching fractions 
    and can be found in the cited references.
  }
  \begin{center}\begin{tabular}{lcccc}
  \hline
                       &\multicolumn{4}{c}{Prediction for the branching fraction [$10^{-6}$] }\\
  Channel              &Ref.~\cite{Rui:2012qq}&Ref.~\cite{Kiselev:2002vz}&Ref.~\cite{Ivanov:2002un} &Ref.~\cite{Ivanov:2006ni} \\
  \hline
  \decay{\Bc}{\Ds\Dzb} & $\phantom{00}2.3\phantom{}\pm0.5\phantom{0}$ & $\phantom{0}4.8\phantom{0}$ & $\phantom{0}1.7\phantom{0}$ & $\phantom{0}2.1\phantom{0}$ \\ 
  \decay{\Bc}{\Ds\Dz}  & $\phantom{00}3.0\phantom{}\pm0.5\phantom{0}$ & $\phantom{0}6.6\phantom{0}$ & $\phantom{0}2.5\phantom{0}$ & $\phantom{0}7.4\phantom{0}$ \\ 
  \decay{\Bc}{\Dp\Dzb} & $32\phantom{}             \pm7\phantom{}$    & $53\phantom{}$              & $32\phantom{}$              & $33\phantom{}$              \\ 
  \decay{\Bc}{\Dp\Dz}  & $\phantom{3}0.10          \pm0.02$           & $\phantom{0}0.32$           & $\phantom{0}0.11$           & $\phantom{0}0.32$           \\ 
  \hline
  \end{tabular}\end{center}
\label{table:bfestimate}
\end{table}

\begin{figure}[t]
\includegraphics[width=8cm]{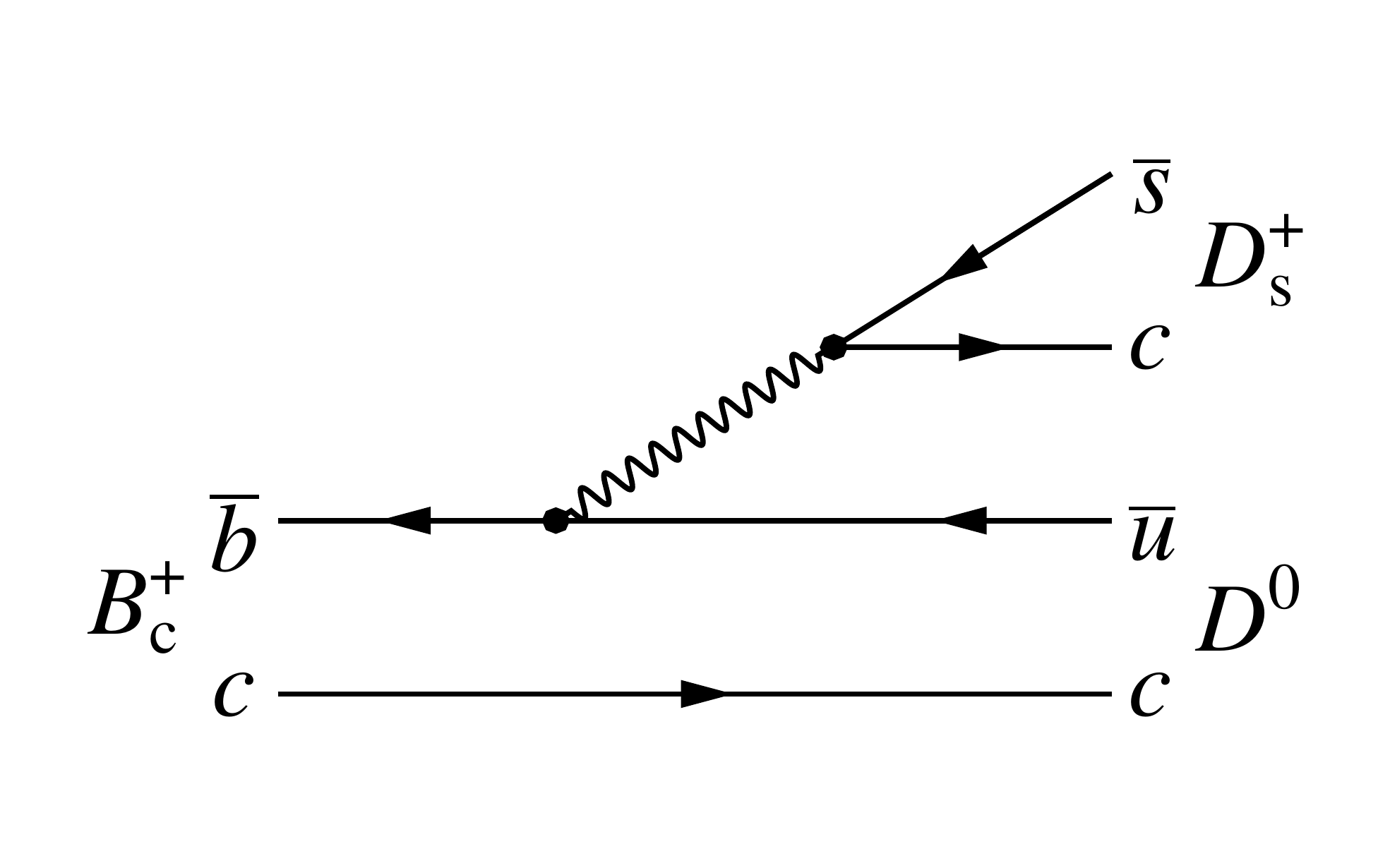}
\includegraphics[width=8cm]{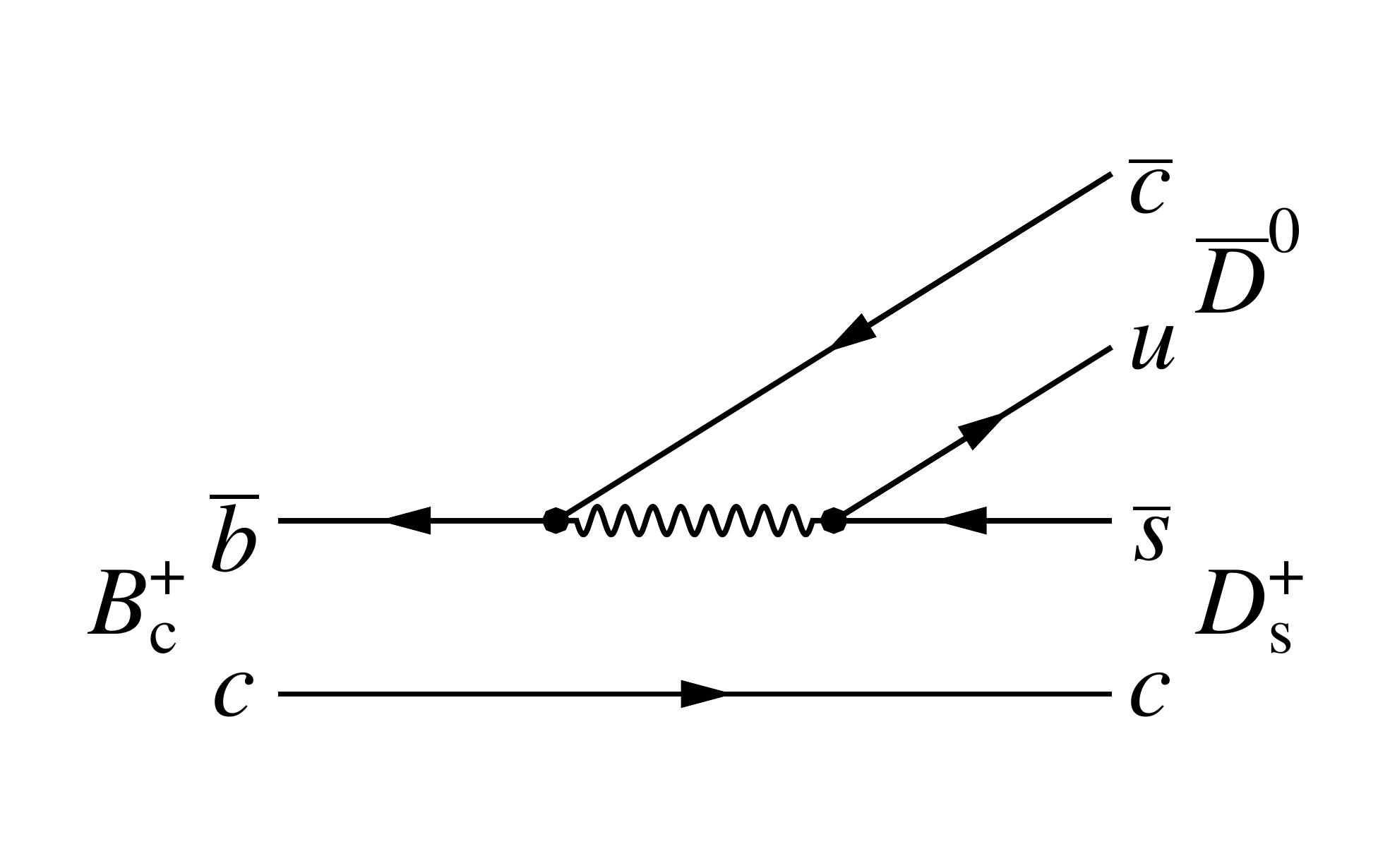}
\caption{Illustration of (left) a colour-favoured \decay{\Bc}{\Ds\Dz} decay, and (right) a colour-suppressed \decay{\Bc}{\Ds\Dzb} decay.}
\label{fig:Diagrams_Bc}
\end{figure}

In the determination of $\gamma$, an advantage of \mbox{\decay{\Bc}{\Dsp\DzorDzb}} decays over \mbox{\decay{\Bp}{\DzorDzb\Kp}} decays is that 
the diagram proportional to \Vcb is colour suppressed,
while the diagram proportional to \Vub is not, as illustrated in Fig.~\ref{fig:Diagrams_Bc}.
This results in a large value for the ratio of amplitudes,
$r_\Bc\equiv|A(\decay{\Bc}{\Dz\Ds})/A(\decay{\Bc}{\Dzb\Ds})|\approx 1$,
and potentially in a large \CP asymmetry for \DzorDzb decays to \CP eigenstates. 
In contrast, in \mbox{\decay{\Bp}{\DzorDzb\Kp}} decays, the small value of
$r_B\equiv|A(\decay{\Bp}{\Dz\Kp})/A(\decay{\Bp}{\Dzb\Kp})|\approx 0.1$
 results in small values of the \CP asymmetry.
However, observing and using \mbox{\decay{\Bc}{\Dsp\DzorDzb}} decays is challenging because of
the small \Bc production cross-section, the short \Bc lifetime, 
the complex final states, and the small branching fractions.

This paper describes a search, performed for the first time,
for twelve \mbox{\decay{\Bc}{\DporDsorstar\DzorDzborstar}} decay channels,
using data collected by the LHCb experiment and corresponding 
to an integrated luminosity of 3.0\invfb, 
of which 1.0\invfb was recorded at a centre-of-mass energy 
$\sqrt{s}=7$\tev and 2.0\invfb at $\sqrt{s}=8$\tev.
Charm mesons are reconstructed in the  
\mbox{\decay{\Dz}{\Km\pip}},
\mbox{\decay{\Dz}{\Km\pip\pim\pip}},
\mbox{\decay{\Dp}{\Km\pip\pip}}, and
\mbox{\decay{\Ds}{\Kp\Km\pip}} decay modes.
For \Bc decays that involve one or more excited charm mesons,
no attempt is made to reconstruct the low-momentum particles from 
the decay of excited charm mesons:
the distribution of the invariant mass of the partially reconstructed final-state
peaks at masses just below the \Bc mass.

The branching fractions, \BF, of \Bc decays to fully reconstructed states are measured
relative to high-yield \decay{\Bp}{\DporDs\Dzb} normalisation modes,
\begin{equation}
\frac{\raisebox{0.3em}{$f_c$}}{f_u}\frac{\BF(\decay{\Bc}{\DporDs\DzorDzb})}{\BF(\decay{\Bp}{\DporDs\Dzb})}=\frac{N(\decay{\Bc}{\DporDs\DzorDzb})}{N(\decay{\Bp}{\DporDs\Dzb})}
\frac{\varepsilon(\decay{\Bp}{\DporDs\Dzb})}{\varepsilon(\decay{\Bc}{\DporDs\DzorDzb})},
\label{eq:bfcalc}
\end{equation}
where \inlfcfu is the ratio of \Bc to \Bp production cross-sections,
$N$ stands for the signal yields, and $\varepsilon$ for the total efficiencies.
For \Bc decays with one excited charm meson, the invariant-mass distributions of \mbox{\decay{\Bc}{\DporDsstar\DzorDzb}} 
and \mbox{\decay{\Bc}{\DporDs\DzorDzbstar}} decays are very similar, and the sum of their branching fractions
is measured, weighted by the branching fraction of the excited charged charm meson 
to a charged charm meson and a low-momentum neutral particle, $\BF(\decay{\DporDsstar}{\DporDs\piz,\gamma})$, 
\begin{multline}
\frac{\raisebox{0.3em}{$f_c$}}{f_u}\frac{  \BF(\decay{\Bc}{\DporDsstar\DzorDzb})\BF(\decay{\DporDsstar}{\DporDs\piz,\gamma}) + \BF(\decay{\Bc}{\DporDs\DzorDzbstar}) }{\BF(\decay{\Bp}{\DporDs\Dzb})}= \\ \frac{N(\decay{\Bc}{\DporDsstar\DzorDzb}) + N(\decay{\Bc}{\DporDs\DzorDzbstar})}{N(\decay{\Bp}{\DporDs\Dzb})} \frac{\varepsilon(\decay{\Bp}{\DporDs\Dzb})}{\varepsilon(\decay{\Bc}{\DporDsstar\DzorDzb, \DporDs\DzorDzbstar})},
\label{eq:bfcalcDst}
\end{multline}
where $\varepsilon(\decay{\Bc}{\DporDsstar\DzorDzb, \DporDs\DzorDzbstar})$ is the average efficiency
of \decay{\Bc}{\DporDsstar\DzorDzb} and \decay{\Bc}{\DporDs\DzorDzbstar} decays.
Branching fractions of \decay{\Bc}{\DporDsstar\DzorDzbstar} are corrected for $\BF(\decay{\DporDsstar}{\DporDs\piz,\gamma})$,
\begin{equation}
\frac{\raisebox{0.3em}{$f_c$}}{f_u}\frac{\BF(\decay{\Bc}{\DporDsstar\DzorDzbstar})}{\BF(\decay{\Bp}{\DporDs\Dzb})}=\frac{1}{\BF(\decay{\DporDsstar}{\DporDs\piz,\gamma})}\frac{N(\decay{\Bc}{\DporDsstar\DzorDzbstar})}{N(\decay{\Bp}{\DporDs\Dzb})}
\frac{\varepsilon(\decay{\Bp}{\DporDs\Dzb})}{\varepsilon(\decay{\Bc}{\DporDsstar\DzorDzbstar})}.
\label{eq:bfcalcDstDst}
\end{equation}

LHCb measurements of $(f_c\BF(\decay{\Bc}{\jpsi\pip}))/(f_u\BF(\decay{\Bp}{\jpsi\Kp}))$
show no significant difference of \inlfcfu between
$\sqrt{s}=7\tev$~\cite{LHCb-PAPER-2012-028} and $\sqrt{s}=8\tev$~\cite{LHCb-PAPER-2014-050}
in the LHCb acceptance.
Predictions for $\BF(\decay{\Bc}{\jpsi\pip})$ range from $6.0\times 10^{-4}$
to $2.9\times 10^{-3}$~\cite{Ebert:2003cn,Chang:1992pt,Qiao:2012hp},
implying a value of \inlfcfu in the range 0.24\%--1.2\%.
Since \mbox{$\BF(\decay{\Bc}{\jpsi\pip})$} is presently not measured,
the results in this paper are expressed as the product of \inlfcfu
and the ratio of \Bc to \Bp branching fractions.

\section{Detector and simulation}
\label{sec:Detector}

The \lhcb detector~\cite{Alves:2008zz,LHCb-DP-2014-002} is a single-arm forward
spectrometer covering the \mbox{pseudorapidity} range $2<\eta <5$,
designed for the study of particles containing \bquark or \cquark
quarks. The detector includes a high-precision tracking system
consisting of a silicon-strip vertex detector surrounding the $pp$
interaction region~\cite{LHCb-DP-2014-001}, a large-area silicon-strip detector located
upstream of a dipole magnet with a bending power of about
$4{\mathrm{\,Tm}}$, and three stations of silicon-strip detectors and straw
drift tubes~\cite{LHCb-DP-2013-003} placed downstream of the magnet.
The polarity of the dipole magnet is reversed periodically throughout data-taking.

The tracking system provides a measurement of the momentum of charged particles with
a relative uncertainty that varies from 0.5\% at low momentum to 1.0\% at 200\gevc.
The minimum distance of a track to a primary $pp$ interaction vertex (PV), the impact parameter (IP), 
is measured with a resolution of $(15+29/\pt)\mum$,
where \pt is the momentum transverse to the beamline expressed in\,\gevc.
Different types of charged hadrons are distinguished using information
from two ring-imaging Cherenkov detectors~\cite{LHCb-DP-2012-003}. 
Photons, electrons and hadrons are identified by a calorimeter system consisting of
scintillating-pad and preshower detectors, an electromagnetic
calorimeter and a hadronic calorimeter. Muons are identified by a
system composed of alternating layers of iron and multiwire
proportional chambers~\cite{LHCb-DP-2012-002}.

The online event selection is performed by a trigger~\cite{LHCb-DP-2012-004}, 
which consists of a hardware stage, based on information from the calorimeter and muon
systems, followed by a software stage, which applies a full event reconstruction.
At the hardware trigger stage, events are required to have a muon with high \pt or a
hadron, photon or electron with high transverse energy in the calorimeters. For hadrons,
the transverse energy threshold is about 3.5\gev.
The software trigger requires a two-, three- or four-track
secondary vertex with a large sum of the transverse momentum of
the tracks and a significant displacement from any PV.
At least one track should have $\pt>1.7\gevc$ and \chisqip with respect to any
PV greater than 16, where \chisqip is defined as the
difference in the vertex-fit \chisq of a given PV reconstructed with and
without the considered particle.
A multivariate algorithm~\cite{BBDT} is used for
the identification of secondary vertices consistent with the decay
of a \bquark hadron.

Simulated events are used for the training of the multivariate selection of the \Bc signals,
for establishing the shape of the invariant-mass distributions of the signals,  and for
determining the relative efficiency between the \Bc signal decays and the \Bp normalisation modes.
In the simulation, $pp$ collisions with \mbox{\decay{\Bp}{\DporDs\Dzb}} decays are generated using
\pythia~\cite{Sjostrand:2007gs,*Sjostrand:2006za} 
with a specific \lhcb configuration~\cite{LHCb-PROC-2010-056}.
For \decay{\Bc}{\DporDs\Dzb} decays, the \bcvegpy~\cite{Chang:2003cq} generator is used.
The simulated \decay{\Bc}{\DporDs\Dzb} sample is also used for training and efficiency calculations
of the \decay{\Bc}{\DporDs\Dz} decay mode.
%The kinematics of \decay{\Bc}{\DporDs\Dz} is expected to be identical to \decay{\Bc}{\DporDs\Dzb} 
%For determining the efficiency of \decay{\Bc}{\DporDs\Dz} decays the \decay{\Bc}{\DporDs\Dzb} sample is used. 
Decays of hadronic particles
are described by \evtgen~\cite{Lange:2001uf}, with final-state
radiation generated using \photos~\cite{Golonka:2005pn}. The
interaction of the generated particles with the detector, and its response,
are implemented using the \geant
toolkit~\cite{Allison:2006ve, *Agostinelli:2002hh} as described in
Ref.~\cite{LHCb-PROC-2011-006}.
Known discrepancies in the simulation are corrected using data-driven methods.

\section{Candidate selection }
\label{sec:selection}

Initially, loose requirements are made to select candidates having both a \DporDs and a \Dz or \Dzb meson.
The charm-meson candidates are constructed by combining two, three or four tracks that are incompatible with  
originating from any reconstructed PV. 
In addition, the tracks must form a high-quality vertex and the scalar sum of their transverse momenta must exceed 1.8\,\gevc.
The pion and kaon candidates
are also required to satisfy loose particle identification (PID) criteria to reduce the contribution to the
selected sample from misidentified particles.
Charm-meson candidates must have an invariant mass within $\pm25\mevcc$ of their known value~\cite{PDG2017}.
Using the same method as in Ref.~\cite{LHCb-PAPER-2014-002},
three-track combinations that are compatible with both \mbox{\decay{\Dp}{\Km\pip\pip}} and \mbox{\decay{\Ds}{\Kp\Km\pip}} decays 
are categorised as a \Ds candidate if the $\Kp\Km$ combination is compatible with the \mbox{\decay{\phi}{\Kp\Km}} decay
or if the $\Kp$ candidate satisfies strict PID criteria, and as a \Dp candidate otherwise.
The two charm mesons are combined into a \BporBc candidate, which is retained if its invariant mass is in the range $4.8-7.0\gevcc$.
The $\DporDs\DzorDzb$ pair must form a good-quality vertex with transverse momentum exceeding 4.0\gevc.
The resulting trajectory of the \BporBc candidate must be consistent with originating from the associated PV,
where the associated PV is the PV with which the \BporBc candidate has the smallest \chisqip.
The reconstructed decay time divided by its uncertainty, $t/\sigma_t$,
of \Dz and \Ds mesons with respect to the \BporBc vertex is required to exceed $-3$,
while that of the longer-lived \Dp meson is required to exceed $+3$.
The tighter decay-time significance criterion on the \Dp eliminates background from \mbox{\decay{\Bp}{\Dzb\pip\pim\pip}} decays 
where the negatively charged pion is misidentified as a kaon.

The invariant-mass resolution of \BporBc decays is significantly improved by applying a kinematic fit~\cite{Hulsbergen:2005pu}
where the masses of the \Dz and the \DporDs candidates are fixed to their known values~\cite{PDG2017},
all particles from the \DporDs, \Dz, or \BporBc decay are constrained to originate from their decay vertex
and the \BporBc is constrained to originate from a PV.

To reduce the combinatorial background, while keeping the efficiency for signal as high as possible,
a multivariate selection based on a boosted decision tree (BDT)~\cite{Breiman,Roe} is employed.
The following variables are used as input for the BDT:
the transverse momentum and the ratio of the likelihood between the kaon and pion PID hypotheses of all final-state particles;
the fit quality of the \BporBc and both charm-meson vertices;
the value of \chisqip of the \BporBc candidate;
the values of $t/\sigma_t$ of the \BporBc and both charm-meson candidates;
the invariant masses of the reconstructed charm-meson candidates; and
the invariant masses of the pairs of opposite-charge tracks from the \DporDs candidate.

Four distinct classifiers are constructed: the BDT training is performed separately
for the \Ds\DzorDzb and \Dp\DzorDzb final states and for the \DKpi and \DKpipipi decay channels.
For a given \Dz final state, the same classifier is used for both \mbox{\decay{\Bc}{\DporDs\Dzb}} and \mbox{\decay{\Bc}{\DporDs\Dz}} decays.
For signal, the BDT is trained using simulated \Bc events,
while for background data in the range $5350<m(\DporDs\DzorDzb)<6200\mevcc$ are used.
Studies indicate that the combinatorial background 
is dominated by non-charm and single-charm candidates, while combinations 
of two real charm mesons contribute less than $5\%$.
To increase the size of the background sample for the BDT training,
the charm mass windows are increased from $\pm25\mevcc$ to $\pm75\mevcc$.

The BDT combines all input variables into a single discriminant.
The optimal value of the cut on this discriminant is determined using 
a procedure based on Ref.~\cite{Punzi:2003bu}, maximising 
$\varepsilon/(\sqrt{N_B}+5/2)$, where $N_B$ is the expected background in a $\pm20\mevcc$ window around the \Bc mass,
and the number 5 is the target significance.
Simulated events are used to estimate the signal efficiency $\varepsilon$.

\section{Data fit}
\label{sec:datafit}

After the selection, a model of the invariant-mass distribution of \decay{\BporBc}{\DporDs\DzorDzb} candidates is fitted to the data.
The model is composed of six components: 
the signals for fully reconstructed \Bp and \Bc decays;
the signal for \Bc decays with one excited charm meson in the final state;
the signal for \Bc decays with two excited charm mesons in the final state;
the background from \mbox{\decay{\Bp}{\Dzb\Kp\Km\pip}} decays;
and the combinatorial background.

Fully reconstructed \Bp and \Bc signals are described by the sum of two Crystal Ball (CB)~\cite{Skwarnicki:1986xj} functions,
with power-law tails proportional to $[m(\DporDs\DzorDzb)-m(\BporBc)]^{-2}$ in opposite directions. 
The peak values of both CB components are constrained to be equal and the other shape parameters
of the CB functions are obtained from a fit to the simulated events.
The peak position of the \Bp signal is a free parameter in the fit to data, while the peak 
position of the \Bc signal is fixed to the world-average measurement~\cite{PDG2017}.
The large \mbox{\decay{\Bp}{\Ds\Dzb}} signal from data is well described by this model.

Models for decays where one or two low-momentum particles from excited charm-meson decays are missing are
implemented as templates, obtained from invariant-mass distributions of simulated data.
For decays with one missing low-momentum particle, both \mbox{\decay{\Bc}{\DporDsstar\DzorDzb}}
and \mbox{\decay{\Bc}{\DporDs\DzorDzbstar}} decays contribute and the template is based on the 
sum of the two decay modes, weighted by the appropriate branching fractions of the excited charm mesons.
For \mbox{\decay{\Bc}{\DporDsstar\DzorDzbstar}} decays, it is assumed that both excited charm mesons are produced unpolarised.

The Cabibbo-favoured \mbox{\decay{\Bp}{\Dzb\Kp\Km\pip}} decay is a background
to the \mbox{\decay{\Bp}{\Ds\Dzb}} channel, though its yield is strongly reduced by the charm-meson mass requirement.
This background is modelled by a single Gaussian function, with the width determined from a sample of simulated decays and the 
normalisation determined from the sidebands of the \Ds mass peak.
The yield of this background is about 40 times smaller than that of the signal, and the shape of the invariant-mass distribution is twice as wide.
The combinatorial background is described by the sum of an exponential function and a constant.

An unbinned extended maximum likelihood fit is used to simultaneously describe the invariant-mass distributions of candidates with 
\DKpi and \DKpipipi decays, resulting in four independent fits to eight invariant mass distributions.
In these fits the background parameters and \Bp yields are free to vary independently,
but the ratio of the \Bc yields for the two \Dz decay modes is constrained
to the corresponding ratio of \Bp yields, corrected for the relative efficiencies.
The total \Bc yield, $N^{\rm tot}_\Bc$, is a free parameter in these fits, 
leading to a \Bc yield in each data sample given by the expressions
\begin{equation}
N^{K\pi}_\Bc=\frac{N_\Bp^{K\pi} \varepsilon_\Bc^{K\pi}/\varepsilon_\Bp^{K\pi} }{ N_\Bp^{K\pi} \varepsilon_\Bc^{K\pi}/\varepsilon_\Bp^{K\pi} + N_\Bp^{K\pi\pi\pi} \varepsilon_\Bc^{K\pi\pi\pi}/\varepsilon_\Bp^{K\pi\pi\pi}  }N^{\rm tot}_\Bc,
\end{equation}
\begin{equation}
N^{K\pi\pi\pi}_\Bc=\frac{N_\Bp^{K\pi\pi\pi} \varepsilon_\Bc^{K\pi\pi\pi}/\varepsilon_\Bp^{K\pi\pi\pi} }{ N_\Bp^{K\pi} \varepsilon_\Bc^{K\pi}/\varepsilon_\Bp^{K\pi} + N_\Bp^{K\pi\pi\pi} \varepsilon_\Bc^{K\pi\pi\pi}/\varepsilon_\Bp^{K\pi\pi\pi}  }N^{\rm tot}_\Bc.
\end{equation}
The relative efficiencies that appear in these expressions,
calculated for simulated events generated in the rapidity range $2.0<y(\BporBc)<4.5$ and with $\pt(\BporBc)>4\gevc$,
are summarised in Table~\ref{table:efficiencies}.

\begin{table}[t]
\caption{Ratio $\varepsilon_{\Bc}/\varepsilon_{\Bp}$ of total efficiencies of \Bc
decays relative to the corresponding fully reconstructed \Bp decays.
The quoted uncertainties are statistical only.}
\begin{center}\begin{tabular}{lcccc}
\hline
                       &\multicolumn{4}{c}{Reconstructed state} \\
                       &\multicolumn{2}{c}{\Ds\DzorDzb with $\Dz\to$} &\multicolumn{2}{c}{\Dp\DzorDzb with $\Dz\to$} \\
Decay channel        &\Km\pip & \Km\pip\pim\pip & \Km\pip & \Km\pip\pim\pip \\
\hline
\decay{\Bc}{\DporDs\DzorDzb}                        & $0.420 \pm 0.005$ & $0.373 \pm 0.009$ & $0.441 \pm 0.007$ & $0.398 \pm 0.010$ \\
\decay{\Bc}{\DporDsstar\DzorDzb,\DporDs\DzorDzbstar}& $0.372 \pm 0.006$ & $0.317 \pm 0.010$ & $0.381 \pm 0.008$ & $0.337 \pm 0.011$ \\
\decay{\Bc}{\DporDsstar\DzorDzbstar}                & $0.339 \pm 0.006$ & $0.278 \pm 0.009$ & $0.342 \pm 0.007$ & $0.297 \pm 0.010$ \\
\hline
\end{tabular}\end{center}
\label{table:efficiencies}
\end{table}

\begin{figure}[tbp]
\includegraphics[width=8.0cm]{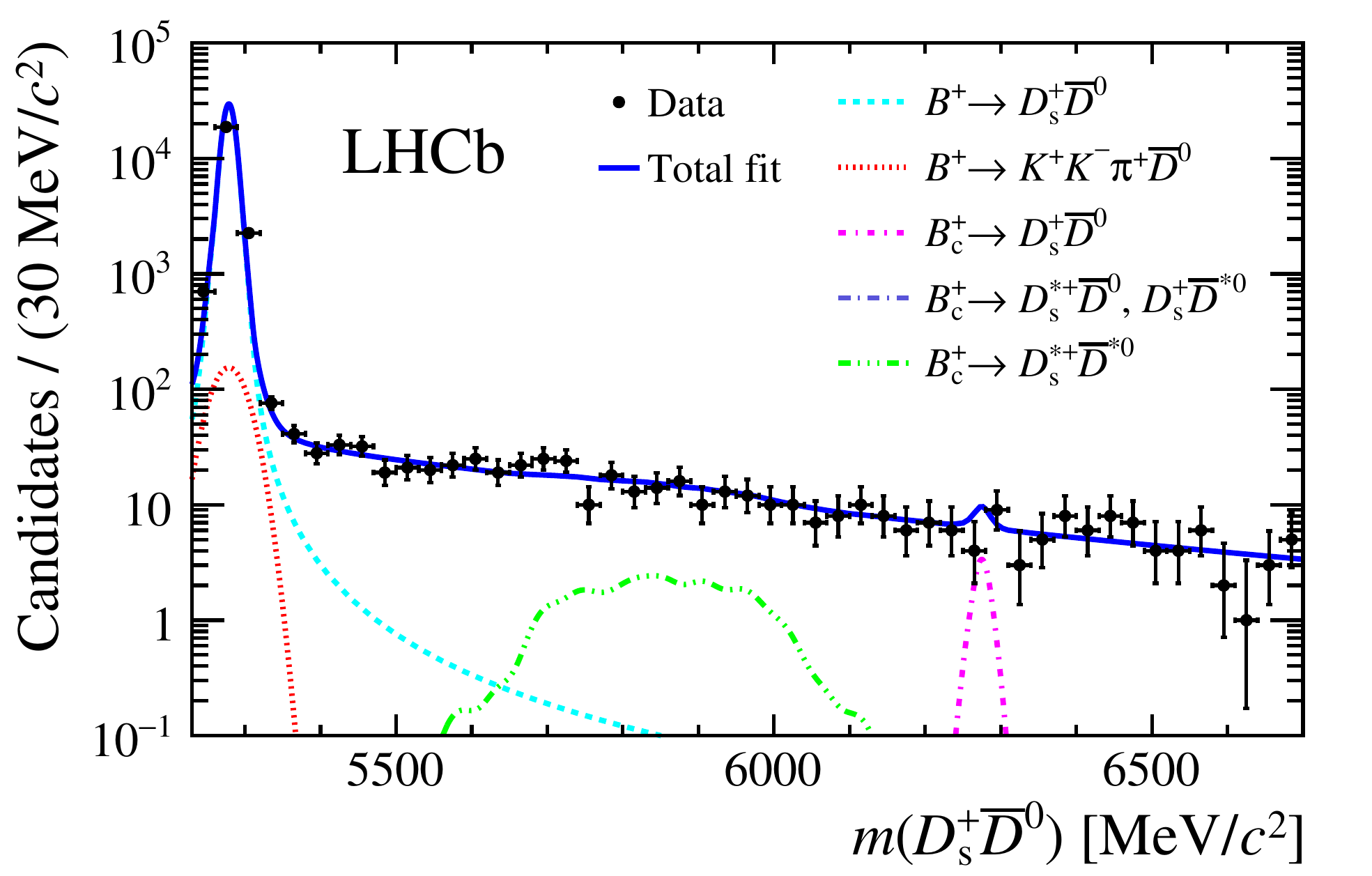}
\includegraphics[width=8.0cm]{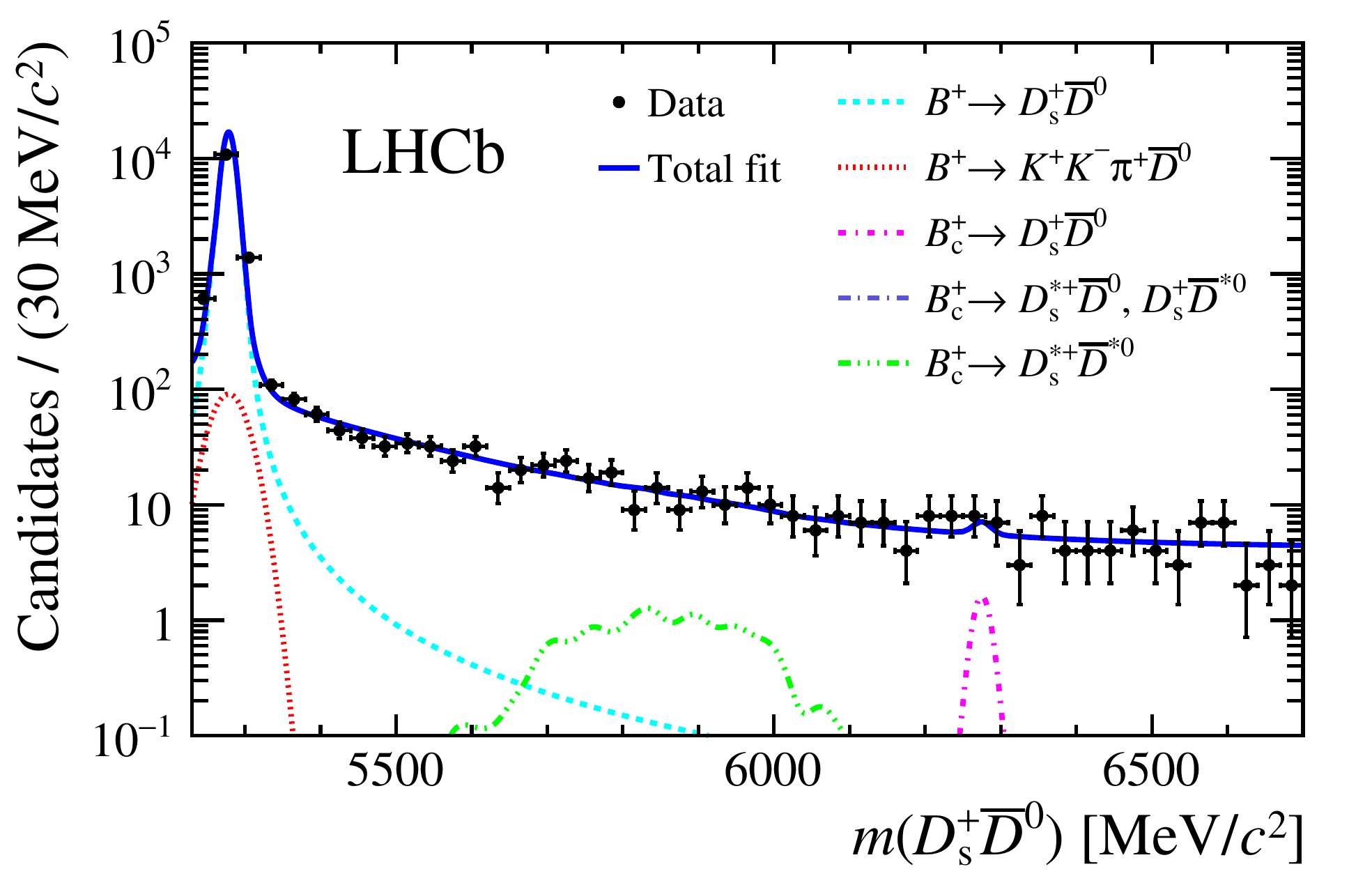}\\
\includegraphics[width=8.0cm]{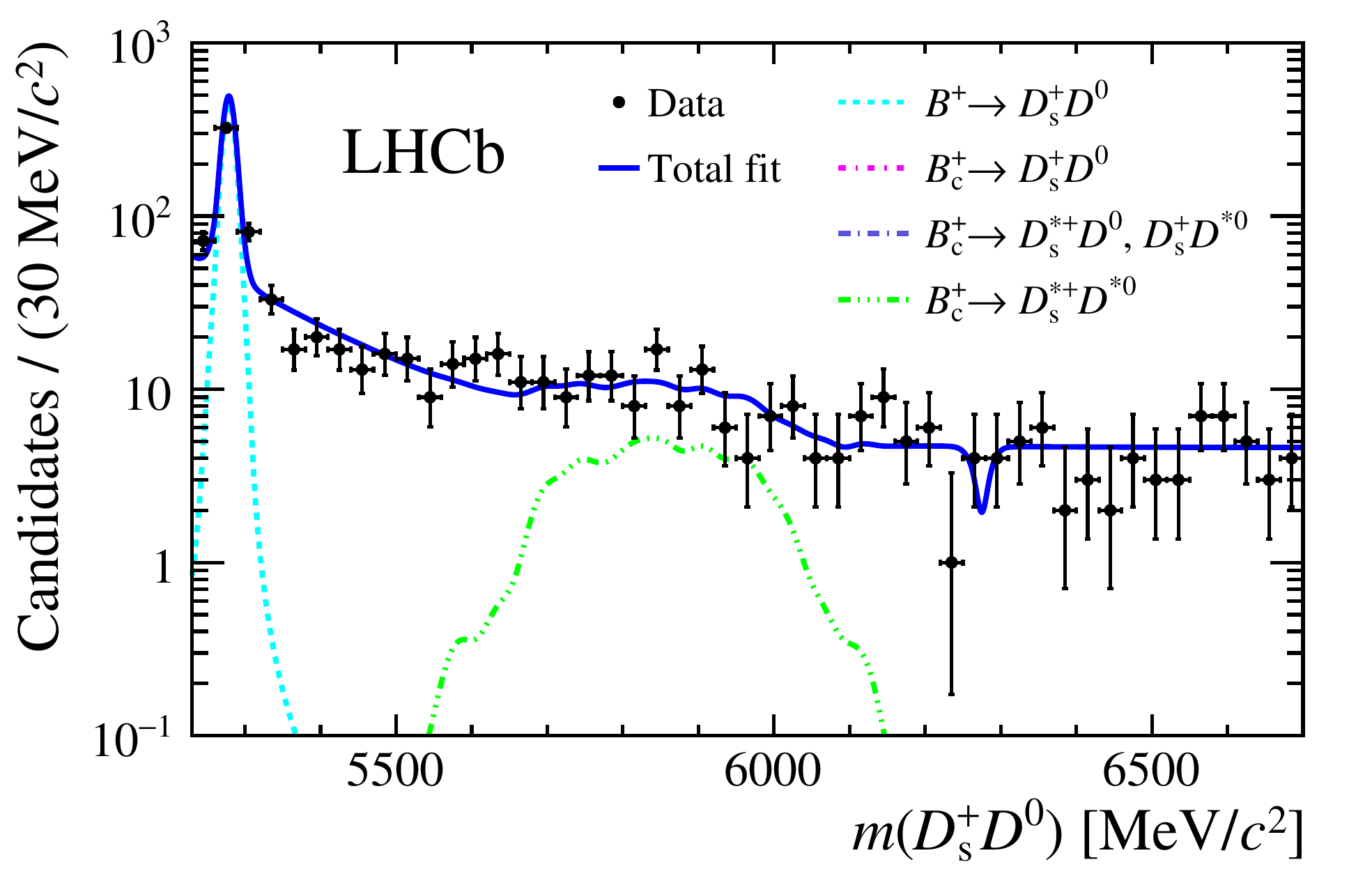}
\includegraphics[width=8.0cm]{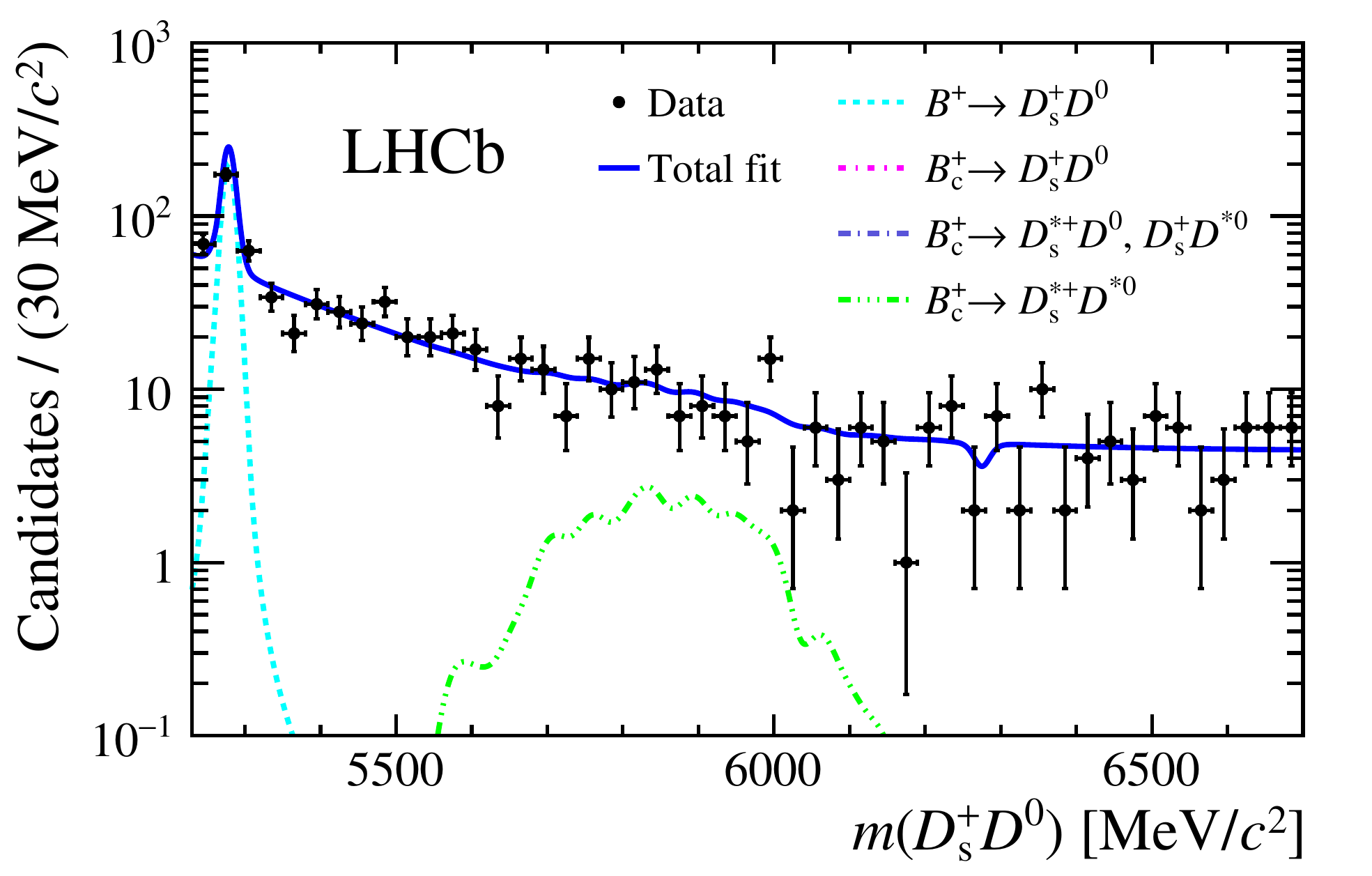}\\
\includegraphics[width=8.0cm]{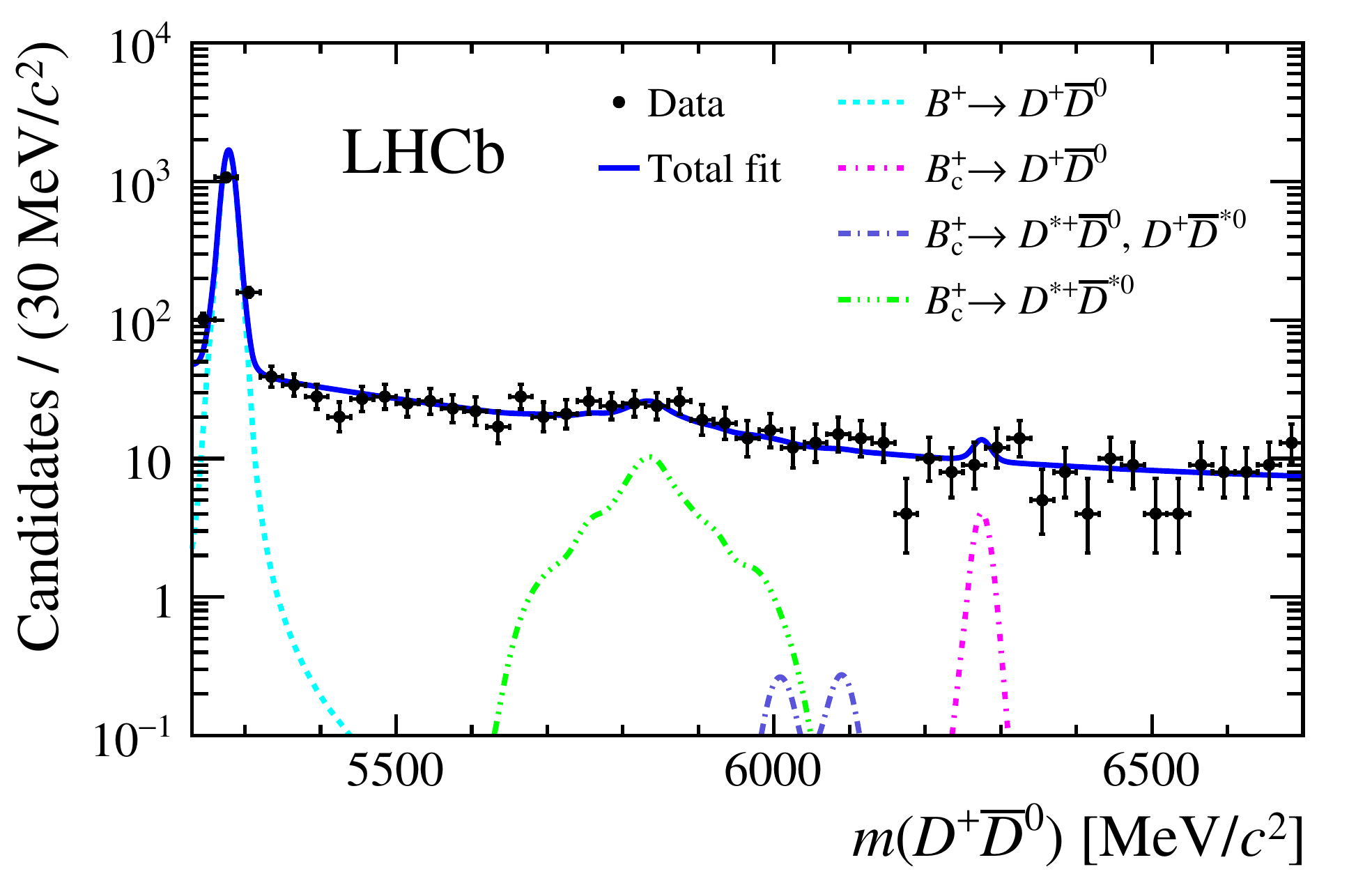}
\includegraphics[width=8.0cm]{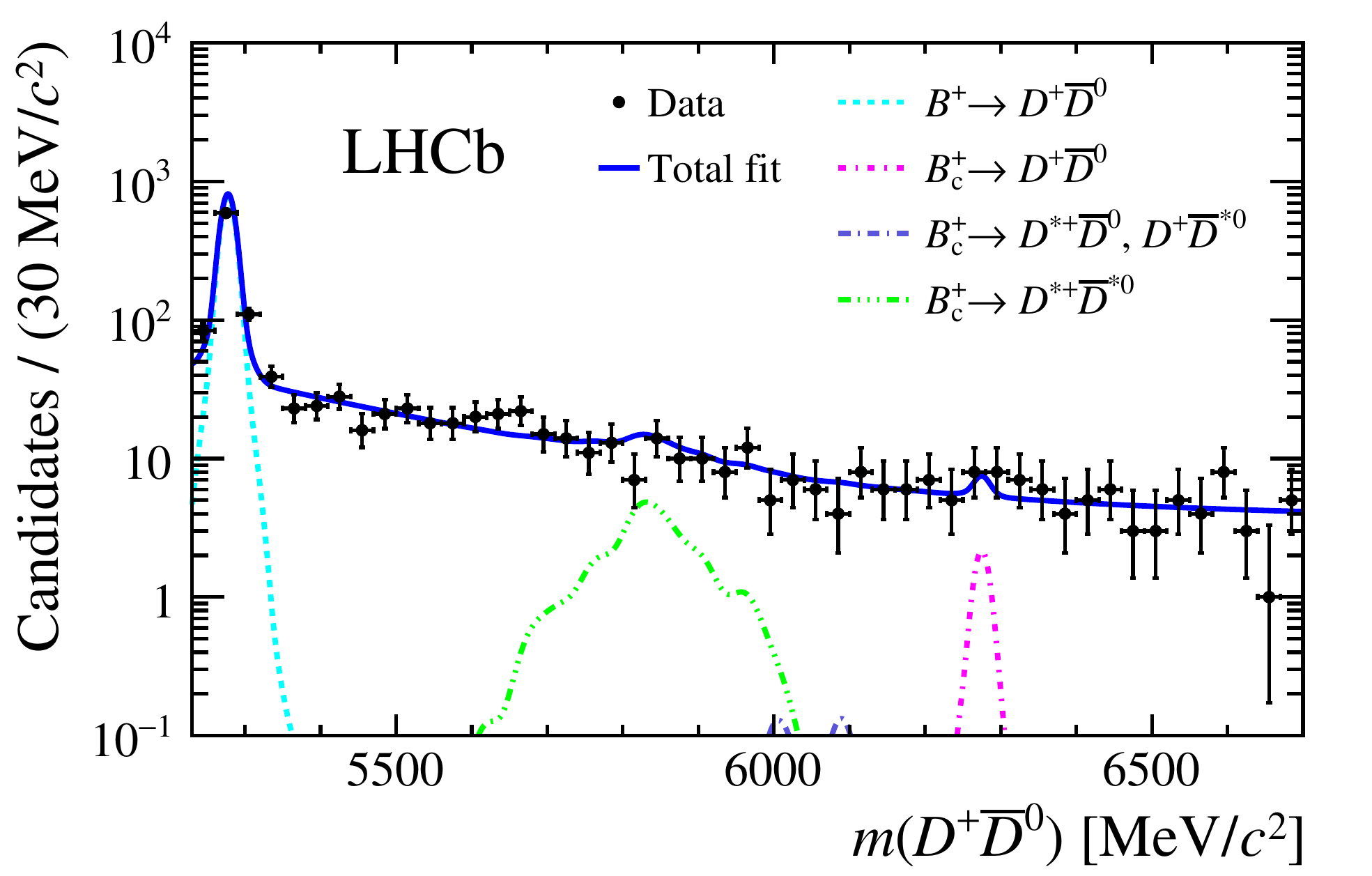}\\
\includegraphics[width=8.0cm]{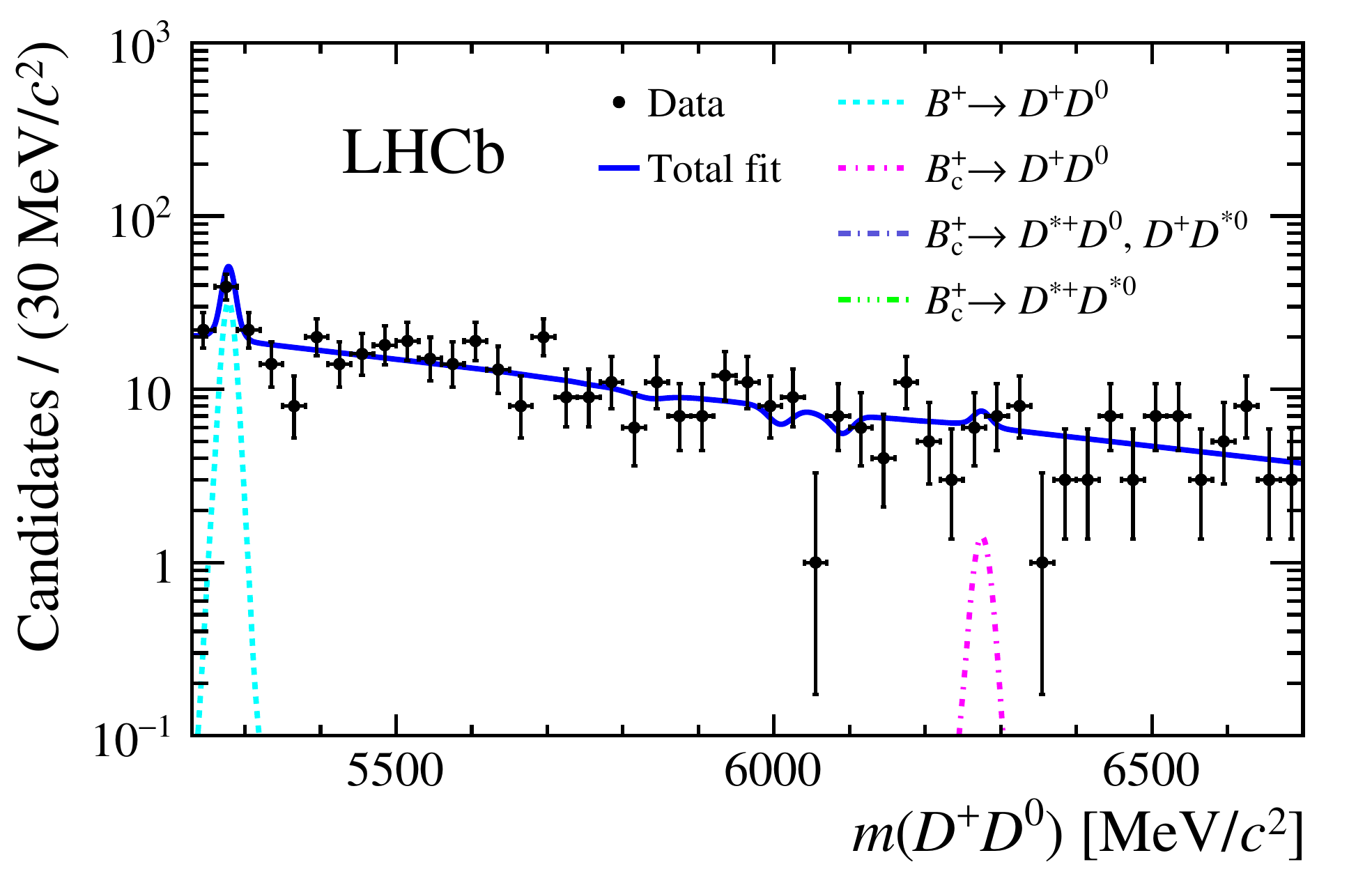}
\includegraphics[width=8.0cm]{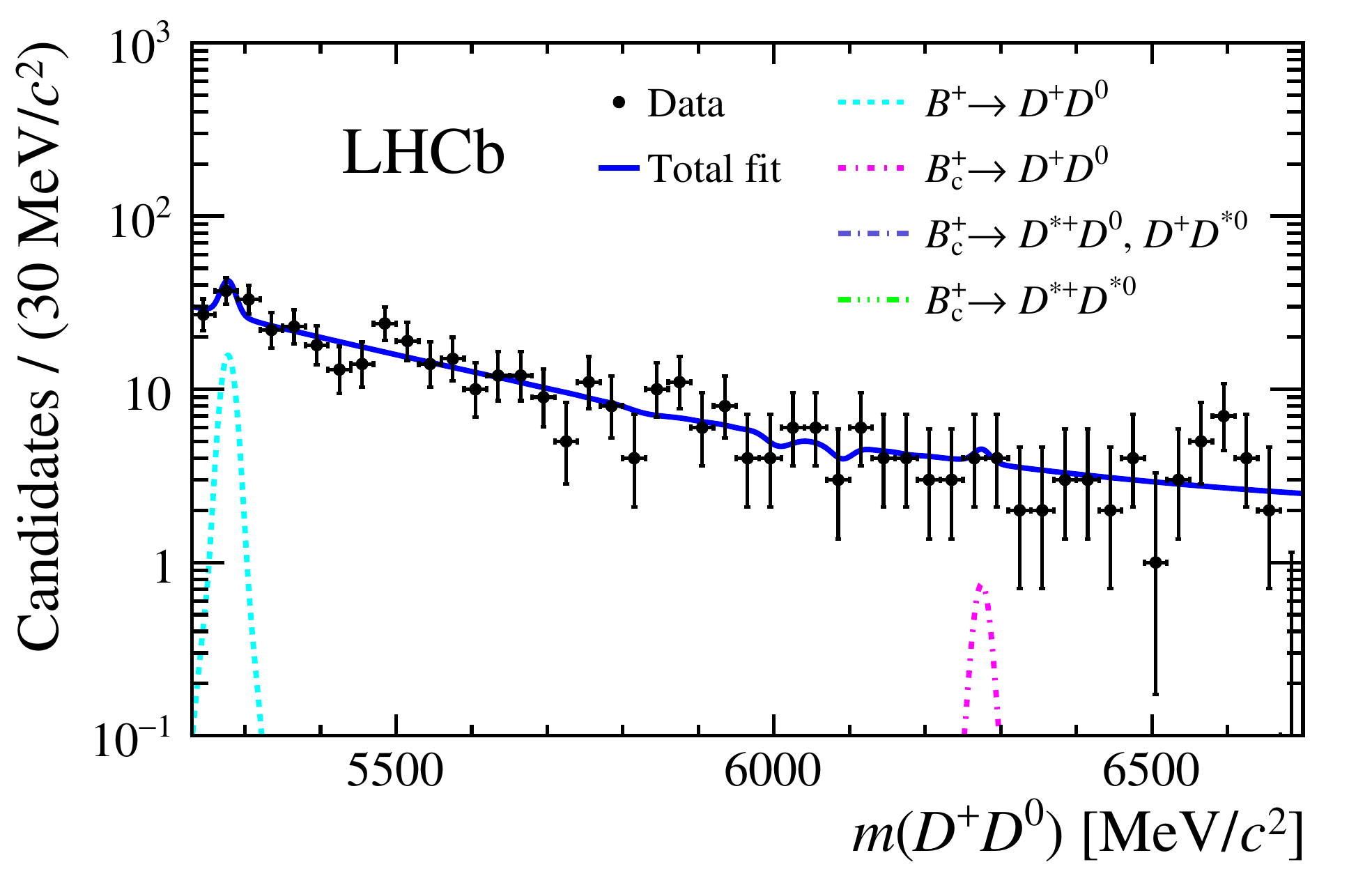}\\
\caption{Fits to the (top row) \Ds\Dzb, (second row) \Ds\Dz, (third row) \Dp\Dzb and (bottom row) \Dp\Dz final states.
For the left plots, the \Dz meson is reconstructed in the $K^-\pi^+$ final state, while the right column 
corresponds to the \mbox{\decay{\Dz}{\Km\pip\pim\pip}} mode.}
\label{fig:fullfit}
\end{figure}

The results of the fits are shown in Fig.~\ref{fig:fullfit},
and the corresponding  signal yields are listed in Table~\ref{table:yields}.
The small peaks at the \Bp mass in the \DporDs\Dz final state are due to \mbox{\decay{\Bp}{\DporDs\Dzb}} decays
either followed by the doubly Cabibbo-suppressed \mbox{\decay{\Dz}{\Kp\pim}} decay
or when both the kaon and pion are misidentified.
No significant \Bc signals are observed;
after taking into account systematic uncertainties, discussed in Sec.~\ref{sec:systematics}, none of the signals
exceeds a significance of two standard deviations,
which is measured as the difference in likelihood when fitting the data with or without signal component in the fit~\cite{Wilks:1938dza}. 

\begin{table}[t]
\caption{Signal yields from the fits of \mbox{\decay{\B}{\DporDs\DzorDzb}} decays.
Samples with \decay{\Dz}{\Km\pip} and \decay{\Dz}{\Km\pip\pim\pip}
are fitted simultaneously. The uncertainties are statistical only.}
\begin{center}
\begin{tabular}{lcccc}
\hline
                       &\multicolumn{4}{c}{Reconstructed state} \\
Decay channel          & \phantom{000}\Ds\Dzb & \phantom{0}\Ds\Dz & \phantom{00}\Dp\Dzb & \phantom{}\Dp\Dz \\
\hline
\decay{\Bp}{\DporDs\Dzb}                & \phantom{$-$}$33\,734\pm\phantom{}187$&\phantom{$-$}$476\pm\phantom{}27$&\phantom{$-$}$1866\pm\phantom{}46$&\phantom{$-$}$37\pm\phantom{}11$\\
\decay{\Bc}{\DporDs\DzorDzb}                  & \phantom{$-0000$}\,$5\pm\phantom{00}5$&\phantom{$00$}$-4\pm\phantom{0}3$&\phantom{$-000$}$6\pm\phantom{0}6$&\phantom{$-0$}$2\pm\phantom{0}4$\\
\decay{\Bc}{\DporDsstar\DzorDzb,\DporDs\DzorDzbstar}& \phantom{$0000$}\,$-1\pm\phantom{0}14$&\phantom{$00$}$-4\pm\phantom{}10$&\phantom{$-000$}$1\pm\phantom{}13$&\phantom{$$}$-10\pm\phantom{0}9$\\
\decay{\Bc}{\DporDsstar\DzorDzbstar}          & \phantom{$-000$}\,$34\pm\phantom{0}28$&\phantom{$-0$}$73\pm\phantom{}19$&\phantom{$-00$}$68\pm\phantom{}23$&\phantom{$0$}$-8\pm\phantom{}14$\\
\hline
\end{tabular}
\end{center}
\label{table:yields}
\end{table}

\boldmath
\section{Systematic uncertainties}
\unboldmath
\label{sec:systematics}

The systematic uncertainties on the \Bc yields are listed in Table~\ref{table:yieldsyst} and described below.
The signal shape parameters for the fully reconstructed modes
are varied according to Gaussian distributions that take into account  the
covariance matrix of the fit to the simulated events,
and evaluating the change in yield and its uncertainty for 1000 variations.
An additional uncertainty is attributed to the signal model
by changing its description from a sum of two CB functions to a sum of two Gaussian functions.
The assumed peak position of the \mbox{\decay{\Bc}{\DporDs\DzorDzb}} signal may differ from the true value.
This is taken into account by varying the \Bc peak position by its uncertainty,
taken as the squared sum of uncertainty on the world-average \Bc
mass ($0.8\mevcc$) and the contribution from the LHCb momentum-scale uncertainty ($0.8\mevcc$)~\cite{LHCb-PAPER-2012-048}.
The signal shape of the decays with one missing low-momentum particle
is based on the assumption \mbox{$\BF(\decay{\Bc}{\DporDsstar\DzorDzb})=\BF(\decay{\Bc}{\DporDs\DzorDzbstar})$}.
Since the \Bc branching fractions are unknown, the signal composition is varied
using \decay{\Bc}{\DporDsstar\DzorDzb} or \decay{\Bc}{\DporDs\DzorDzbstar} only 
and the largest difference is taken as the systematic uncertainty. 
As the polarisation of excited charm mesons in \mbox{\decay{\Bc}{\DporDsstar\DzorDzbstar}} decays is unknown,
the signal shapes are varied between fully longitudinal and fully transverse polarisations,
and the largest yield difference with the unpolarised decay model is taken as the uncertainty.
To evaluate the uncertainty in the choice of the shape of the combinatorial background,
an alternative fit is applied using an exponential function to model the background.
To evaluate eventual biases of the \Bc yields in the fit, 
pseudoexperiments are generated where the candidates in the signal window are replaced by 
the expected distribution using only background. The yields are corrected for this bias 
and the attributed uncertainty is the squared sum of the bias and its statistical uncertainty.

\begin{table}[t]
\caption{
Systematic uncertainties on the \Bc yields, for the combined fit to both the \DKpi and the \DKpipipi decay channels.
The total systematic uncertainty is calculated as the quadratic sum of the individual components.}
\begin{center}\begin{tabular}{lcccc}
\hline
          &\multicolumn{4}{c}{Reconstructed state} \\
Source             & \Ds\Dzb       &  \Ds\Dz       &  \Dp\Dzb      &  \Dp\Dz     \\
\hline
\hline
\decay{\Bc}{\DporDs\DzorDzb} &           &               &               &             \\
\hline
Signal shape       &     0.25      &       0.28    &        0.31   &   0.13      \\
Signal model       &     0.40      &       0.34    &        0.61   &   0.44      \\
\Bc mass           &     0.64      &       0.62    &        0.79   &   0.51      \\
Background model   &     1.12      &       1.75    &        1.88   &   0.56      \\
Fit bias           &     0.70      &       1.28    &        0.27   &   0.19      \\
\hline
Total              &     1.54      &       2.30    &        2.17   &   0.91      \\
\hline
\hline
\decay{\Bc}{\DporDsstar\DzorDzb,\DporDs\DzorDzbstar} & &       &               &             \\
\hline
Signal composition &\phantom{0}7.6 &\phantom{0}5.5 &\phantom{0}7.1 &   5.7       \\
Background model   &          11.9 &          17.5 &          16.4 &   4.5       \\
Fit bias           &\phantom{0}5.5 &\phantom{0}9.4 &\phantom{0}3.9 &   1.3       \\
\hline
Total              &     15.2      &      20.6     &       18.3    &   7.4       \\
\hline
\hline
\decay{\Bc}{\DporDsstar\DzorDzbstar} &   &               &               &             \\
\hline
Polarisation      &     23         &      14       &\phantom{0}9   &\phantom{0}5 \\
Background model  &     43         &      98       &       37      &\phantom{0}9 \\
Fit bias          &     10         &\phantom{0}7   & \phantom{0}8  &\phantom{0}1 \\
\hline
Total             &     49         &      99       &        39     &   10        \\
\hline
  \end{tabular}\end{center}
\label{table:yieldsyst}
\end{table}

Systematic uncertainties that affect the normalisation are listed in Table~\ref{table:multsyst} and are described below.
The limited size of the simulated signal samples affects the normalisation as well as 
the statistical uncertainties of the \Bp yields.
The systematic uncertainties of the \Bp yields are evaluated by varying the signal shape according to the
covariance matrix of the fit to simulated data and by changing the signal model to the sum of two Gaussian functions.
The \Bp yield is also affected by uncertainties on the background, which are evaluated by  
changing the background shape to an exponential function and by varying the single-charm background by 100\% of its yield.
The impact on the efficiency ratio of the uncertainty on the \Bc lifetime  
is evaluated by changing its lifetime by one standard deviation.
Imperfections in the rescaling of the PID variables~\cite{LHCb-DP-2012-003} are quantified by considering 
the efficiency ratio with and without PID corrections and assigning the difference as a systematic uncertainty.
The \DKpipipi decay has a complicated substructure, but was simulated according to a phase-space model.
The systematic uncertainty is taken as the quadratic sum of the differences
in efficiency ratio when the simulated events are weighted
to reproduce the $\pip\pim$, $\Km\pip$, $\Km\pip\pim$ and $\pip\pim\pip$ invariant-mass
distributions observed in data.
The difference in efficiency when applying the model variations for \Bc decays
with one or two excited charm mesons in the final state
is taken into account as a systematic uncertainty.
The determinations of the \mbox{\decay{\Bc}{\Dstarp\DzorDzbstar}} branching fraction ratios
are corrected for \mbox{$\BF(\decay{\Dstarp}{\Dp\piz,\gamma})=(32.3\pm0.5)\%$}~\cite{PDG2017},
as is indicated in Eq.~\ref{eq:bfcalcDstDst},
and the corresponding uncertainty is assigned as a systematic uncertainty.

\begin{table}[t]
\caption{
Systematic uncertainties, in \%, on the normalisation of the \Bc branching fraction determination.
The total systematic uncertainty is calculated as the quadratic sum of the individual components.}
\begin{center}\begin{tabular}{llcccc}
\hline
       & &\multicolumn{4}{c}{Reconstructed state}\\
       & & \multicolumn{2}{c}{\Ds\DzorDzb, with $\Dz\rightarrow$} & \multicolumn{2}{c}{\Dp\DzorDzb, with $\Dz\rightarrow$} \\
Channel&  Source                                      & \Km\pip &  \Km\pip\pim\pip & \Km\pip &  \Km\pip\pim\pip\\
\hline
Common & \Bp stat.                                          &  0.7  &  0.9  &  3.1  &  4.3 \\
       & \Bp signal shape                                   &  0.0  &  0.0  &  0.0  &  0.3 \\
       & \Bp signal model                                   &  0.1  &  0.2  &  0.1  &  0.3 \\
       & Background model                                   &  0.0  &  0.6  &  1.6  &  1.3 \\
       & \decay{\Bp}{\Dzb\Kp\Km\pip}                        &  1.4  &  1.4  &  ---  &  --- \\
       & \Bc lifetime                                       &  1.5  &  1.5  &  1.5  &  1.5 \\
       & PID                                                &  2.4  &  0.9  &  1.2  &  3.2 \\
       & \Dz model                                          &  ---  &  1.1  &  ---  &  0.7 \\
\hline
\decay{\Bc}{\DporDs\DzorDzb} & Simulation stat.                   &  1.2  &  2.4  &  1.6  &  2.5 \\
       & Total                                              &  3.5  &  3.6  &  4.3  &  6.3 \\
\hline
\decay{\Bc}{\DporDsstar\DzorDzb,\DporDs\DzorDzbstar} & Simulation stat. &  1.7  &  3.3  &  2.0  &  3.3 \\
       & Signal composition                                 &  1.0  &  0.8  &  0.7  &  2.6 \\
       & Total                                              &  3.8  &  4.3  &  4.5  &  7.1 \\
\hline                                                                                 
\decay{\Bc}{\DporDsstar\DzorDzbstar} & Simulation stat.           &  1.7  &  3.4  &  2.0  &  3.3 \\
       & Polarisation                                       &  1.5  &  0.4  &  1.4  &  1.3 \\
       & \BF(\decay{\Dstarp}{\Dp\piz,\gamma})               &  ---  &  ---  &  1.5  &  1.5 \\
       & Total                                              &  3.9  &  4.4  &  4.9  &  6.9 \\
\hline
\end{tabular}\end{center}
\label{table:multsyst}
\end{table}

\boldmath
\section{Results and conclusion}
\unboldmath
\label{sec:results}

To determine the branching fraction ratios, fits to data are performed
where the free parameters are not the individual yields, but
correspond to the left-hand-side terms of Eqs.~\ref{eq:bfcalc}--\ref{eq:bfcalcDstDst}.
In these fits, the systematic uncertainties are taken into account as Gaussian constraints.

The measured branching fraction ratios for the fully reconstructed \Bc decays are listed below.
Quoted in brackets are the corresponding upper limits calculated
at 90\%\,(95\%) confidence level with the asymptotic CL$_s$ method~\cite{CLS},
\begin{align*} 
\fcfu\frac{\BF(\decay{\Bc}{\Ds\Dzb})}{\BF(\decay{\Bp}{\Ds\Dzb})}                                                     &= (\phantom{-}3.0\pm3.7)\times 10^{-4}\;[< 0.9\,(1.1)\times 10^{-3}],\\
\fcfu\frac{\BF(\decay{\Bc}{\Ds\Dz})}{\BF(\decay{\Bp}{\Ds\Dzb})}                                                      &= (          -3.8\pm2.6)\times 10^{-4}\;[< 3.7\,(4.7)\times 10^{-4}],\\
\fcfu\frac{\BF(\decay{\Bc}{\Dp\Dzb})}{\BF(\decay{\Bp}{\Dp\Dzb})}                                                     &= (\phantom{-}8.0\pm7.5)\times 10^{-3}\;[< 1.9\,(2.2)\times 10^{-2}],\\
\fcfu\frac{\BF(\decay{\Bc}{\Dp\Dz})}{\BF(\decay{\Bp}{\Dp\Dzb})}                                                      &= (\phantom{-}2.9\pm5.3)\times 10^{-3}\;[< 1.2\,(1.4)\times 10^{-2}].\\
\end{align*}
For \Bc decays with one excited charm meson, the results are
\begin{align*} 
\fcfu\frac{\BF(\decay{\Bc}{\Dss\Dzb})+\BF(\decay{\Bc}{\Ds\Dstarzb})}{\BF(\decay{\Bp}{\Ds\Dzb})}                      &= (          -0.1\pm1.5)\times 10^{-3}\;[< 2.8\,(3.4)\times 10^{-3}],\\
\fcfu\frac{\BF(\decay{\Bc}{\Dss\Dz})+\BF(\decay{\Bc}{\Ds\Dstarz})}{\BF(\decay{\Bp}{\Ds\Dzb})}                        &= (          -0.3\pm1.9)\times 10^{-3}\;[< 3.0\,(3.6)\times 10^{-3}],\\
\fcfu\frac{\BF(\decay{\Bc}{(\Dstarp\to\Dp\piz,\gamma)\Dzb})+\BF(\decay{\Bc}{\Dp\Dstarzb})}{\BF(\decay{\Bp}{\Dp\Dzb})}&= (\phantom{-}0.2\pm3.2)\times 10^{-2}\;[< 5.5\,(6.6)\times 10^{-2}],\\
\fcfu\frac{\BF(\decay{\Bc}{(\Dstarp\to\Dp\piz,\gamma)\Dz})+\BF(\decay{\Bc}{\Dp\Dstarz})}{\BF(\decay{\Bp}{\Dp\Dzb})}  &= (          -1.5\pm1.7)\times 10^{-2}\;[< 2.2\,(2.8)\times 10^{-2}].\\
\end{align*}
For \Bc decays with two excited charm mesons, the measurements give
\begin{align*} 
\fcfu\frac{\BF(\decay{\Bc}{\Dss\Dstarzb})}{\BF(\decay{\Bp}{\Ds\Dzb})}                                                &= (\phantom{-}3.2\pm4.3)\times 10^{-3}\;[< 1.1\,(1.3)\times 10^{-2}],\\
\fcfu\frac{\BF(\decay{\Bc}{\Dss\Dstarz})}{\BF(\decay{\Bp}{\Ds\Dzb})}                                                 &= (\phantom{-}7.0\pm9.2)\times 10^{-3}\;[< 2.0\,(2.4)\times 10^{-2}],\\
\fcfu\frac{\BF(\decay{\Bc}{\Dstarp\Dstarzb})}{\BF(\decay{\Bp}{\Dp\Dzb})}                                             &= (\phantom{-}3.4\pm2.3)\times 10^{-1}\;[< 6.5\,(7.3)\times 10^{-1}],\\
\fcfu\frac{\BF(\decay{\Bc}{\Dstarp\Dstarz})}{\BF(\decay{\Bp}{\Dp\Dzb})}                                              &= (          -4.1\pm9.1)\times 10^{-2}\;[< 1.3\,(1.6)\times 10^{-1}].
\end{align*}

The presented limits are consistent with the theoretical expectations: 
assuming a value of $\inlfcfu=1.2\%$, the branching fraction ratio limits
give $\BF(\decay{\Bc}{\Dp\Dzb})<6.0\,(7.0)\times 10^{-4}$ at 90\%\,(95\%) confidence level,
well above the values shown in Table~\ref{table:bfestimate}.

\section*{Acknowledgements}
%
% These Acknowledgements valid from 6-Dec-2017
%
\noindent We express our gratitude to our colleagues in the CERN
accelerator departments for the excellent performance of the LHC. We
thank the technical and administrative staff at the LHCb
institutes. We acknowledge support from CERN and from the national
agencies: CAPES, CNPq, FAPERJ and FINEP (Brazil); MOST and NSFC
(China); CNRS/IN2P3 (France); BMBF, DFG and MPG (Germany); INFN
(Italy); NWO (The Netherlands); MNiSW and NCN (Poland); MEN/IFA
(Romania); MinES and FASO (Russia); MinECo (Spain); SNSF and SER
(Switzerland); NASU (Ukraine); STFC (United Kingdom); NSF (USA).  We
acknowledge the computing resources that are provided by CERN, IN2P3
(France), KIT and DESY (Germany), INFN (Italy), SURF (The
Netherlands), PIC (Spain), GridPP (United Kingdom), RRCKI and Yandex
LLC (Russia), CSCS (Switzerland), IFIN-HH (Romania), CBPF (Brazil),
PL-GRID (Poland) and OSC (USA). We are indebted to the communities
behind the multiple open-source software packages on which we depend.
Individual groups or members have received support from AvH Foundation
(Germany), EPLANET, Marie Sk\l{}odowska-Curie Actions and ERC
(European Union), ANR, Labex P2IO and OCEVU, and R\'{e}gion
Auvergne-Rh\^{o}ne-Alpes (France), RFBR, RSF and Yandex LLC (Russia),
GVA, XuntaGal and GENCAT (Spain), Herchel Smith Fund, the Royal
Society, the English-Speaking Union and the Leverhulme Trust (United
Kingdom).

% This should be taken out in the final paper
%\input{supplementary-app}

\clearpage

\addcontentsline{toc}{section}{References}
\setboolean{inbibliography}{true}
\bibliographystyle{LHCb}
\bibliography{main,LHCb-PAPER,LHCb-CONF,LHCb-DP,LHCb-TDR}

\newpage

% Author List ----------------------------
\centerline{\large\bf LHCb collaboration}
\begin{flushleft}
\small
R.~Aaij$^{40}$,
B.~Adeva$^{39}$,
M.~Adinolfi$^{48}$,
Z.~Ajaltouni$^{5}$,
S.~Akar$^{59}$,
J.~Albrecht$^{10}$,
F.~Alessio$^{40}$,
M.~Alexander$^{53}$,
A.~Alfonso~Albero$^{38}$,
S.~Ali$^{43}$
% ...

{\footnotesize \it
% ...
$ ^{5}$Clermont Universit{\'e}, Universit{\'e} Blaise Pascal, CNRS/IN2P3, LPC, Clermont-Ferrand, France\\
% ...
$ ^{10}$Fakult{\"a}t Physik, Technische Universit{\"a}t Dortmund, Dortmund, Germany\\
% ...
$ ^{38}$ICCUB, Universitat de Barcelona, Barcelona, Spain\\
$ ^{39}$Instituto Galego de F{\'\i}sica de Altas Enerx{\'\i}as (IGFAE), Universidade de Santiago de Compostela, Santiago de Compostela, Spain\\
$ ^{40}$European Organization for Nuclear Research (CERN), Geneva, Switzerland\\
% ...
$ ^{43}$Nikhef National Institute for Subatomic Physics, Amsterdam, The Netherlands\\
% ...
$ ^{48}$H.H. Wills Physics Laboratory, University of Bristol, Bristol, United Kingdom\\
% ...
$ ^{53}$School of Physics and Astronomy, University of Glasgow, Glasgow, United Kingdom\\
% ...
$ ^{59}$University of Cincinnati, Cincinnati, OH, United States\\
% ...
}
\end{flushleft}

\end{document}